\documentclass[%
 reprint,
superscriptaddress,
 amsmath,amssymb, 
 aps,
prb,
floatfix,
]{revtex4-2}

\usepackage[normalem]{ulem}
\usepackage{xcolor}
\usepackage{graphicx}
\usepackage{dcolumn}
\usepackage{bm}
\usepackage{hyperref}

\usepackage{braket}
\usepackage{multirow}

\newcommand{\MS}[1]{\text{MS}_{#1}}
\newcommand{\YM}[1]{\text{YM}_{#1}}
\newcommand{\ED}[1]{\text{ED}_{#1}}

\begin{document}

\preprint{APS/123-QED}

\title{Energy minimization of paired composite fermion wave functions in the spherical geometry}

\author{Greg J. Henderson}
\affiliation{Rudolf Peierls Centre for Theoretical Physics, Parks Road, Oxford OX1 3PU, United Kingdom}
\author{Gunnar M{\"o}ller}
\affiliation{University of Kent, Canterbury CT2 7NZ, United Kingdom}
\author{Steven H. Simon}
\affiliation{Rudolf Peierls Centre for Theoretical Physics, Parks Road, Oxford OX1 3PU, United Kingdom}

\date{\today}

\begin{abstract}
    We perform the energy minimization of the paired composite fermion (CF) wave functions, proposed by M{\"o}ller and Simon (MS) [PRB 77, 075319 (2008)] and extended by Yutushui and Mross (YM) [PRB 102, 195153 (2020)], where the energy is minimized by varying the CF pairing function, in the case of an approximate model of the Coulomb interaction in the second Landau level for pairing channels $\ell = -1, 3, 1$ which are expected to be in the Pfaffian, anti-Pfaffian and particle-hole symmetric (PH) Pfaffian phases respectively.  It is found that the energy of the $\ell = -1$ MS wave function can be reduced substantially below that of the Moore-Read wave function at small system sizes, however, in the $\ell = 3$ case the energy cannot be reduced much below that of the YM trial wavefunction. Nonetheless, both our optimized and unoptimized wavefunctions with $\ell=-1,3$ extrapolate to roughly the same energy per particle in the thermodynamic limit. For the $\ell = 1$ case, the optimization makes no qualitative difference and these PH-Pfaffian wave functions are still energetically unfavourable. The effective CF pairing is analyzed in the resulting wave functions, where the effective pairing for the $\ell = -1, 3$ channels is found to be well approximated by a weak-pairing BCS ansatz and the $\ell = 1$ wave functions show no sign of emergent CF pairing.
\end{abstract}

\maketitle


\section{Introduction} \label{Intro}
Well over 30 years since its discovery \cite{willett_observation_1987}, the precise nature of the fractional quantum Hall effect in the half-filled second Landau level in GaAs heterostructures still remains elusive. The three leading candidate phases of matter, all of which could potentially host non-abelian anyons, are the Pfaffian \cite{Moore1991}, the anti-Pfaffian \cite{storni_fractional_2010, levin_particle-hole_2007, lee_particle-hole_2007} and the particle-hole symmetric Pfaffian (PH-Pfaffian) \cite{Son2015}.  All three of these phases can be understood as paired composite fermions (CFs) at pairing channels $\ell = -1$ \cite{greiter_paired_1991, park_possibility_1998, Read2000}, $\ell = 3$ \cite{barkeshli2015particle} and $\ell = 1$ \cite{barkeshli2015particle, zucker_stabilization_2016} respectively. Numerical studies have pointed toward either the Pfaffian or anti-Pfaffian as being the most likely to occur in this setting \cite{morf_transition_1998, rezayi_incompressible_2000, feiguin_density_2008, Rezayi2011, simon_landau_2013, Zaletel2015, rezayi_landau_2017, Rezayi2021}. However, experimental measurements of the heat conductivity along the edges of these systems \cite{banerjee_observation_2018} and measurements of noise along interfaces of these systems with other quantum Hall states \cite{Dutta2021} are more consistent with the PH-Pfaffian state. We are thus left with an apparent contradiction.

Several proposals have been made in an attempt to resolve this contradiction with one possibility being that the PH-Pfaffian phase could be induced by disorder \cite{mross_theory_2018, lian_theory_2018, wang_topological_2018} and another being the incomplete thermal equilibration of the edge modes affecting the measurements of heat conductivity along the edge \cite{Simon2018, ma_partial_2019, Simon2019, Asasi2020}. While these lines of inquiry might be physically relevant, neither approach has yet provided a fully acceptable explanation of all the experimental facts. 

At the same time, it is worth noting that the numerical work that has been performed on the 5/2 state is necessarily limited to finite-sized systems.   Exact diagonalization, and even DMRG, are limited to fairly modest system sizes. While it seems a bit unlikely that the results obtained at small sizes (particularly that the PH-Pfaffian is unfavored) will change at larger system sizes, given the enduring conflict with experimental observation, it is worth exploring this possibility.
To fully resolve this issue it would then be useful to be able to perform numerical simulations at larger system sizes, than previously studied, with accurate numerically tractable trial wave functions, which is what we shall pursue in the current work.

Whilst the Moore-Read (MR) wave function, which is a representative of the Pfaffian phase, is numerically tractable to large system sizes, it does have a relatively low overlap with the exact Coulomb ground state, in the second Landau level (LL), compared with other trial wave functions for other states in the lowest Landau level (LLL) \cite{morf_transition_1998, rezayi_incompressible_2000}. Furthermore, both the representative trial wave functions for the anti-Pfaffian, taken to be the particle-hole conjugate of the MR wave functions denoted $\overline{\text{MR}}$, and the PH-Pfaffian, proposed in Ref. \cite{zucker_stabilization_2016}, are only numerically tractable at small system sizes. There is also mounting evidence that the current representative trial wave function for the PH-Pfaffian does not in fact represent a gapped phase of matter \cite{balram_parton_2018, mishmash_numerical_2018, pakrouski_approximate_2021}.

Motivated in part by this poor overlap of the MR wave function with the exact ground state, M{\"o}ller and Simon (MS) proposed a numerically tractable trial wave function for the Pfaffian phase, denoted $\MS{-1}$, which can be interpreted as a wave function of paired CFs at pairing channel $\ell = -1$, where the pairing function can be varied \cite{Moller2008}. By varying the pairing function so as to minimize the energy of the wave function MS where then able to obtain a more accurate approximation to the exact ground state. This MS construction was later extended by Yutushui and Mross (YM) to CF paring channels $\ell = 3$ (denoted $\MS{3}$) and $\ell = 1$ (denoted $\MS{1}$) which are expected to be in the anti-Pfaffian and PH-Pfaffian phases respectively. YM studied these wave functions with fixed pairing functions (i.e. with the variational parameters fixed), where they found the $\ell = 3$ wave function to be an accurate representative of the $\overline{\text{MR}}$ wave function and found the $\ell = 1$ wave function to show no sign of emergent paring of CFs in the density-density correlation function. In their work, YM actually proposed two versions of these wave functions: the single-particle projected and so-called ``pair-projected'' wave functions. Here we shall only be interested in the single-particle projected wave functions as these are more numerically tractable. We denote the single-particle projected fixed parameter wave functions proposed by YM at pairing channel $\ell$ by $\YM{\ell}$.

A comparison of the energetics of the $\MS{\ell}$ versus the $\YM{\ell}$ wave functions have not yet been performed, and it is not known if the $\MS{\ell}$ wave functions with their energy minimized can offer significantly better approximations to the corresponding ground states at $\ell = 1,3$. In particular, minimising the energy of the $\MS{1}$ wave functions could, in principle, produce PH-Pfaffian trial wave functions with energies significantly closer to the $\ell = -1, 3$ trial wave functions. Performing such optimizations comes with a practical challenge in that they must be done in an efficient way in order to access larger system sizes.

There has also been some recent interest in understanding if the effective CF pairing in these phases can be approximated by some weak pairing BCS-type description, where the BCS gap parameter can be estimated \cite{Sharma2021}. As well as offering some physical insight into these phases of matter, it also has some practical use, by allowing, for example, to observe when the system is transitioning from CF pairing to the CF Fermi liquid \cite{Moller2008}. The precise BCS weak pairing description has only so far been studied for the $\ell = -1$ pairing \cite{Moller2008,Sharma2021} and it is not known if such a description is accurate for the $\ell = 3, 1$ cases.

In this work, we will show how the energy of the $\MS{\ell}$ wave functions can be minimized at larger system sizes, in the spherical geometry, where we will present the results of this optimization for the case of an approximate model of the Coulomb interaction in the second LL in the absence of LL mixing. At small system sizes, we find, in agreement with Ref. \cite{Moller2008}, that the energies of the $\MS{-1}$ wave functions can be reduced substantially below the corresponding energy of the MR wave function, with the optimized $\MS{-1}$ wave functions showing a much-improved overlap with the corresponding exact ground state. It is further observed that the optimized $\MS{3}$ wave functions offer considerably less energy reduction over the corresponding zero-parameter trial wave functions in comparison to the $\MS{-1}$ wave functions, although the optimized $\MS{3}$ still show a noticeable improvement in the overlap with the corresponding exact ground state compared with the $\YM{3}$ wave function. Furthermore, the amount by which the energy of the $\MS{1}$ wave functions can be minimized compared with the $\YM{1}$ wave functions is negligible where we find both wave functions to be energetically unfavourable. We demonstrate that the effective CF pairing in the optimized $\MS{\ell}$ and $\YM{\ell}$ wave functions, for $\ell = -1, 3$, can be well approximated by a weak pairing BCS type description, where, for finite size systems, the $\MS{\ell}$ wave functions show stronger pairing than the corresponding $\YM{\ell}$ wave functions. However,  whether we consider $\YM{\ell}$ or $\MS{\ell}$ with $\ell=-1,3$ or MR wave functions, within our numerical error these all extrapolate to roughly the same energy per particle in the thermodynamic limit. Finally, further pathologies of the $\MS{1}$ wave functions are found where no evidence of emergent pairing between the CF (at this pairing channel) is found even after some optimization.

The $\MS{\ell}$ and $\YM{\ell}$ wave functions are introduced in Sec. \ref{Sec:MSWF}. Then in Sec. \ref{Sec:effInteractions} we present the approximate LLL model of the Coulomb interaction in the first excited LL. Finally, in Sec. \ref{Sec:results} the results of these energy minimizations are presented in the case of the approximate model of the Coulomb interaction in the second LL, where it is also shown how the effective weak-pairing BCS description can be extracted from these wave functions. The optimization algorithm used in this work is detailed in Appendix \ref{Sec:optimization}.

Throughout this work, we will assume the $\nu = 5/2$ system to be such that the LLL is completely full for the spin-up and spin-down electron orbitals and that the second LL is at half-filling with the electrons being spin polarized. This second LL system is then mapped to the LLL in the usual way (see Sec. \ref{Sec:effInteractions}). Note that we will not include any Landau level mixing in the calculations presented in this work.

\section{Paired CF wave functions on the sphere} \label{Sec:MSWF}
In the BCS theory of superconductivity \cite{bardeen_theory_1957} one typically starts with the \textit{mean-field} Hamiltonian for a system of fermions, which we will take to be in two spatial dimensions, which takes the form
\begin{equation}
    H_{\text{BCS}} = \sum_{\mathbf{k}} \bigg [ \varepsilon_\mathbf{k} c^\dagger_\mathbf{k} c_\mathbf{k} + \frac{1}{2} ( \overline{\Delta}_{\mathbf{k}}c_{-\mathbf{k}}c_\mathbf{k} + \Delta_\mathbf{k} c^\dagger_\mathbf{k} c^\dagger_{-\mathbf{k}} ) \bigg ]
\end{equation}
and is assumed to be a reasonable approximation to the actual Hamiltonian of the system at low energies, where $\Delta_\mathbf{k}$ is known as the \textit{gap function} and $\varepsilon_\mathbf{k}$ is the kinetic energy relative to the Fermi level of a single particle state labelled by $\mathbf{k}$ (i.e. $\varepsilon_\mathbf{k} = E_\mathbf{k} - \mu$ with $\mu$ the chemical potential). The unnormalised ground state of this Hamiltonian is given by,
\begin{equation}
    \ket{\Psi_{\text{BCS}}} = \exp \bigg ( \frac{1}{2} \sum_\mathbf{k} g_\mathbf{k} c^\dagger_{-\mathbf{k}}c^\dagger_\mathbf{k} \bigg ) \ket{0}
\end{equation}
where $g_\mathbf{k} = (\varepsilon_\mathbf{k} - \sqrt{\varepsilon_{\mathbf{k}}^2 + |\Delta_\mathbf{k}|^2})/\overline{\Delta}_\mathbf{k}$. Note that for this case of spinless fermions $g_\mathbf{k}$ must be an odd function of $\mathbf{k}$, $g_{-\mathbf{k}} = - g_\mathbf{k}$. In real-space this can be expressed as
\begin{equation}
    \ket{\Psi_{\text{BCS}}} = \exp \bigg ( \frac{1}{2} \int d^2 \mathbf{r}_2 d^2\mathbf{r}_2 g(\mathbf{r}_1 - \mathbf{r}_2) c^\dagger(\mathbf{r}_1)c^\dagger(\mathbf{r}_2) \bigg ) \ket{0}
\end{equation}
$\ket{\Psi_{\text{BCS}}}$ is physically interpreted as a \textit{paired state}, where particles near the Fermi level become bound into pairs with $g(\mathbf{r})$ being referred to as the \textit{pairing function}. 

The average occupation of the orbital labelled by $\mathbf{k}$, $n_\mathbf{k}$, can be expressed as $n_\mathbf{k} = \braket{\Psi_{\text{BCS}} | \Psi_{\text{BCS}}}^{-1} \bra{\Psi_{\text{BCS}}} c^\dagger_\mathbf{k}c_\mathbf{k}\ket{\Psi_{\text{BCS}}} = \braket{\Psi_{\text{BCS}} | \Psi_{\text{BCS}}}^{-1} (g_\mathbf{k}-g_{-\mathbf{k}}) \bra{\Psi_{\text{BCS}}}  \partial_{g_\mathbf{k}}\ket{\Psi_{\text{BCS}}} = \\ \braket{\Psi_{\text{BCS}} | \Psi_{\text{BCS}}}^{-1} 2(g_\mathbf{k}) \bra{\Psi_{\text{BCS}}}  \partial_{g_\mathbf{k}}\ket{\Psi_{\text{BCS}}} = |g_\mathbf{k}|^2/(1+|g_\mathbf{k}|^2)$. This can be expressed in terms of $\varepsilon_\mathbf{k}$ and $\Delta_\mathbf{k}$ as $n_\mathbf{k} = \frac{1}{2} \big ( 1 - \frac{\varepsilon_\mathbf{k}}{\sqrt{\varepsilon_\mathbf{k}^2 + |\Delta_\mathbf{k}|^2}} \big )$. 

For a rotationally symmetric microscopic Hamiltonian, the gap function is some eigenstate of the rotation operator. In two dimensions this would mean that under a rotation by angle $\vartheta$ the gap function transforms as $\Delta_\mathbf{k} \rightarrow e^{i \ell \vartheta}\Delta_\mathbf{k}$, where $\ell$ is known as the \textit{pairing channel}. For spinless fermion cases we are considering here $\Delta_\mathbf{k}$ is an odd function of $\mathbf{k}$, as $g_\mathbf{k}$ is, and a typical ansatz for $\Delta_\mathbf{k}$ is $\Delta_\mathbf{k} = \Delta |k|e^{i\ell\theta}$, where $\theta$ is the angle from the x-axis in $\mathbf{k}$-space. For $|\mathbf{r}|\gg \Delta/\mu$ the pairing function corresponding to this gap function is $g(\mathbf{r}) \propto \frac{e^{i\ell \theta}}{|\mathbf{r}|}$ (see Appendix A of Ref. \cite{Yutushui2020}). When this long-distance form of $g$ occurs the fermions are said to be in a weak-pairing phase\cite{Read2000}. 

This BCS mean-field wave function can be used to create a trial wave function for a fixed number of fermions $N$, by projecting it to the space of states with $N$ particles, with the projector written as $P_N$. We then define $\ket{\Psi_N} \equiv P_N \ket{\Psi_{\text{BCS}}}$. The wave function of $\ket{\Psi_N}$ is given by $\Psi_N(\mathbf{r}_1,\mathbf{r}_2, \dots , \mathbf{r}_N) = \text{Pf} [ g(\mathbf{r}_i - \mathbf{r}_j) ]$. Note that as $[c^\dagger_\mathbf{k}c_\mathbf{k}, P_N] = 0$, we still have the property that $c^\dagger_\mathbf{k}c_\mathbf{k} \ket{\Psi_N} = 2g_\mathbf{k}\partial_{g_\mathbf{k}}\ket{\Psi_N}$. Hence, the average orbital occupations are given by $n_\mathbf{k} = \braket{\Psi_N | \Psi_N}^{-1} 2g_\mathbf{k} \bra{\Psi_N} \partial_{g_\mathbf{k}}\ket{\Psi_N}$. 

Now we move to a system of $N$ spinless fermions moving on a sphere with $N_\phi = 2Q = 2(N - 1 + q)$ flux quanta passing through its surface, where $q$ is on the order of one (i.e. does not scale with $N$) and it is presumed that the magnetic field is strong enough that the fermions are confined to the LLL. Thus, in the thermodynamic limit, this system is at filling fraction $\nu = \lim_{N\rightarrow \infty} \frac{N}{N_\phi} = \frac{1}{2}$. Let us then assume they form CFs with each fermion being bound to 2 wave function vortices. The effective flux that the CFs experience is then $N^*_\phi = 2q$. As the effective magnetic field is negligible in the thermodynamic limit, if the effective interaction between the CFs is weakly attractive we then expect them to form some weakly paired BCS state. Let $u_i, v_i$ be the spinor coordinates for $i^{\text{th}}$ fermion. On the sphere, the flux attaching Jastrow factor is $\prod_{i<j}^N(u_iv_j-v_iu_j)^2$. From standard CF theory we then write the ``ideal'' trial wave function as $\Psi = P_{LLL} \text{Pf}[g(\mathbf{r}_i - \mathbf{r}_j)]\prod_{i<j}^N(u_iv_j-v_iu_j)^2$, where, by imposing rotational invariance, the pairing function takes the general form $g(\mathbf{r}_i - \mathbf{r}_j) = \sum_{lm} (-1)^{m+|q|} g_l Y_{qlm}(\mathbf{\Omega}_i) Y_{ql(-m)}(\mathbf{\Omega}_j)$ for some unspecified $g_l \in \mathbb{C}$ with $Y_{qlm}(\Omega)$ being the monopole harmonics \cite{Wu1976}. In principle, one can then extract the pairing physics of the CFs by finding the $g_l$ that minimizes the energy of this wave function. 

This is, however, numerically intractable for $N \gtrapprox 10$ due to the projection to the LLL. To create a numerically tractable trial wave function MS proposed using the Jain-Kamila \cite{Jain1997} procedure where we can produce a LLL wave function simply by replacing the single-particle orbitals $Y_{qlm}(\mathbf{\Omega}_i)$ by the corresponding CF ``orbitals'' defined by,
\begin{equation}
    \Tilde{Y}_{qlm}(\mathbf{\Omega}_i) \equiv J_i^{-1} [P_{LLL}Y_{qlm}(\mathbf{\Omega}_i)J_i]
\end{equation}
where $J_i = \prod_{j\neq i}^N (u_iv_j - v_iu_j)$ and $P_{LLL}$ projects particle $i$ to the LLL with magnetic flux $2q + N -1$. The resulting family of pairing wave functions then defines MS paired CF wave functions \cite{Moller2008},
\begin{equation}
    \begin{split}
        \Psi_{\text{MS}} = & \text{Pf} \bigg [ \sum_{lm} (-1)^{m+|q|} g_l \Tilde{Y}_{qlm}(\mathbf{\Omega}_i) \Tilde{Y}_{ql(-m)}(\mathbf{\Omega}_j) \bigg ] \\
        & \times \prod_{i<j}^N(u_iv_j-v_iu_j)^2
    \end{split}
\end{equation}
In particular, we denote the MS family of wave functions at effective flux $q$ by $\text{MS}_{2q}$.
One can then vary the $g_l$ to minimize the energy. The YM wave function at effective flux $2q$, denoted $\YM{2q}$, is defined to be the $\MS{2q}$ wave function with $g_l = \frac{1}{2l+1}$. Note that as the interaction potential $V(\mathbf{r})$ is real-valued, we must have that $\bra{\Psi_{\text{MS}}} V \ket{\Psi_{\text{MS}}}$ is invariant under time reversal where we simply replace the wave function by its complex conjugate. Furthermore, by expressing $\bra{\Psi_{\text{MS}}} V \ket{\Psi_{\text{MS}}}$ as an integral one can perform a change of variable where we change the azimuthal angle $\phi \rightarrow -\phi$ which is the same as the transformation $u \rightarrow u^*$ and $v \rightarrow v^*$. Combining these two transformations we must have that $\bra{\Psi_{\text{MS}}} V \ket{\Psi_{\text{MS}}}$ is invariant under $g_l \rightarrow g_l^*$. Assuming that the minimum energy solution is unique up to multiplying all $g_l$ by the same complex number, it then follows that the minimum energy wave function can be expressed with $g_l$ all being real numbers. We will take $g_l \in \mathbb{R}$ from now on.

From the usual CF theory, we expect that this can be physically interpreted as being analogous to a BCS state, of the CFs, of the form
\begin{equation}
    \ket{\Psi_{\text{BCS}}} = \exp \bigg ( \frac{1}{2} \sum_{lm}g_l(-1)^{m+|q|} c^\dagger_{qlm} c^\dagger_{ql(-m)} \bigg )\ket{0}
\end{equation}

where $c^\dagger_{qlm}$ is the creation operator for the orbital $Y_{qlm}(\mathbf{\Omega})$. The pairing wave function of this BCS state is $g(\mathbf{r}_i-\mathbf{r}_j) = \sum_{lm}g_l(-1)^{m+|q|} Y_{qlm}(\mathbf{\Omega}_i)Y_{ql(-m)}(\mathbf{\Omega}_j)$. This expansion of the pair wave functions was shown to describe pairing in the $\ell=2q$ channel \cite{mollerPhDThesis2006, Moller2008BilayerPRL, Moller2009BilayerPRB}, as follows from the properties of the Dirac monopole harmonics on the sphere \cite{Wu1977} by considering relations for the complex conjugation of these functions (see Eq.~A5 in \cite{Moller2009BilayerPRB}).
Alternatively, using Eq. B4 of Ref. \cite{Yutushui2020} the pair wave function can be expressed as $g(\mathbf{r}_i-\mathbf{r}_j) = \sum_l \frac{g_l(-1)^{|q|-q}(2l+1)}{4\pi} (u_iv_j-v_iu_j)^{2q}P_{l-q}^{(2q,0)}(\cos \theta)$, where $\theta$ is the angle between particle $i$ and particle $j$ and $P^{(2q,0)}_n$ are the Jacobi polynomials, where it should be noted that if $q<0$ then one should replace $(u,v) \rightarrow (u^*, v^*)$. Here, the Jastrow factor $(u_iv_j-v_iu_j)^{2q}$ allows one to read off the pairing channel explicitly, as the remainder of the expression is real, confirming the result $\ell = 2q$\footnote{This can be seen using stereographic coordinates with $z_i = u_i/v_i$ which then gives $(u_iv_j-v_iu_j)^2q = v_i^{2q}v_j^{2q}(z_i-z_j)^{2q}$. Hence, as we bring particle $i$ around particle $j$ the phase of the wave function will wind $2q$ times in the anti-clockwise direction.}. Also from Eq. B6b of Ref. \cite{Yutushui2020} we have $\frac{(u_iv_j-v_iu_j)^{q-1/2}}{(u^*_iv^*_j-v_i^*u_j^*)^{q+1/2}} = \sum_{l=q}^\infty (-1)^{q+m}\frac{4\pi}{2l+l} Y_{qlm}(\mathbf{\Omega}_i)Y_{ql(-m)}(\mathbf{\Omega}_j)$, where again negative $q$ requires complex conjugation. This pairing function scales with the distance between particle $i$ and particle $j$ as $g \sim 1/r$. Thus the analogous weak pairing on the sphere is given by $g_l \sim \frac{1}{2l+1}$ at small $l$, the form assumed by YM \cite{Yutushui2020}. Now let $\varepsilon_l = \frac{\hbar^2}{2m_{CF}R^2}(l(l+1) - l_F(l_F+1))$ where $R$ is the radius of the sphere, $m_{CF}$ the the CF effective mass and $l_F$ is the Fermi ``shell''. We can express $g_l = (\varepsilon_l - \sqrt{\varepsilon^2_l + \Delta_l^2})/\Delta_l^*$ where $\Delta_l$ is the gap function on the sphere. We can then make the analogous weak pairing ansatz for the gap function $\Delta_l = \Delta \frac{l + 1/2}{R}$. By matching the kinetic energy for $l \gg 1$ one can see that the correspondence between $l$ and the wave vector $k$ is $k = \frac{l}{R}$ for large $l$. This then reproduces the previous ansatz for the gap function in terms of $k$ for large $l$, $\Delta_l \approx \Delta k$. By symmetry, the occupation of all orbitals in the $l$ shell must be the same $n_l$, which can then be expressed as $n_l = \frac{2g_l}{2l+1} \braket{\Psi_{\text{BCS}} | \Psi_{\text{BCS}}}^{-1} \bra{\Psi_{\text{BCS}}} \partial_{g_l} \ket{\Psi_{\text{BCS}}}$. For the weak pairing ansatz, this gives,
\begin{equation} \label{Eq:nAnsatz}
    n_l = \frac{1}{2}\bigg ( 1 - \frac{\varepsilon_l}{\sqrt{\varepsilon_l^2 + \big ( \frac{\Delta (l + 1/2)}{R} \big )^2 }} \bigg )
\end{equation}

Whilst this is perhaps an appealing way to physically interpret the MS wave function, it is not obvious if the CF orbitals are ``normalised'' in such a way that the $g_l$ of the MS wave function should be the same as the $g_l$ of the effective BCS state. As discussed by MS in Ref. \cite{Moller2008}, one can get around this issue by defining the effective CF occupations as $n^{\text{CF}}_l \equiv \frac{2g_l}{2l+1} \braket{\Psi_{\text{MS}}| \Psi_{\text{MS}}}^{-1} \bra{\Psi_{\text{MS}}} \partial_{g_l} \ket{\Psi_{\text{MS}}}$. This can also be expressed as,
\begin{equation} \label{Eq:CFn}
    n^{\text{CF}}_l = \frac{2g_l}{2l+1} \bigg \langle  \frac{\partial_{g_l} \Psi_{\text{MS}}}{\Psi_{\text{MS}}} \bigg \rangle
\end{equation}
where the expectation value is taken with respect to $\ket{\Psi_{\text{MS}}}$. From their definition, these clearly do not depend on the normalisation of $g_l$. These effective occupation probabilities for the minimum energy $g_l$ were empirically found by MS to behave as the occupation probabilities of an actual fermion system. We will then use these $n^{\text{CF}}_l$ in this work to relate optimized MS wave functions to their effective CF-BCS description.

As the pairing channel is given by $\ell = 2q$, we expect that the MS wave functions $\MS{-1}$, $\MS{3}$ and $\MS{1}$ to be in the Pfaffian, anti-Pfaffian and PH-Pfaffian phases respectively. In what follows we will also use yet another variational wave function for the Pfaffian phase which we will denote MR* and can be expressed as,
\begin{equation}
    \begin{split}
        \Psi_{\text{MR*}} = & \text{Pf} \bigg [ \frac{1}{u_iv_j - v_iu_j} \\
        & + \sum_{lm} (-1)^{m+1/2} g_l \Tilde{Y}_{(-\frac{1}{2})lm}(\mathbf{\Omega}_i) \Tilde{Y}_{(-\frac{1}{2})l(-m)}(\mathbf{\Omega}_j) \bigg ] \\
        & \times \prod_{i<j}^N (u_iv_j - v_iu_j)^2 \\
    \end{split}
\end{equation}
Roughly speaking, the MR* wave function allows us to ``perturb'' around the MR wave function, where if we take the $g_l$ to zero we recover exactly the MR wave function. By varying the $g_l$ parameters, one may expect that an MR* wave function can always be found with lower energy than the corresponding MR wave function (at the same system size). We will thus use the MR* wave function as a benchmark for the optimization of the standard $\MS{-1}$ wave function. 

\section{Model interaction} \label{Sec:effInteractions}
As is standard in the FQHE literature we model a system of electrons confined to the first excited Landau level (LL), and without LL mixing, by an effective description of electrons in the LLL. In the spherical geometry, where the actual system of interest has magnetic flux $2Q$ passing through the sphere's surface, the effective interaction $V^{\text{eff}}$ for the LLL system is defined by,
\begin{equation}
    \begin{split}
        & \bra{Q+1,Q+1, m_1, m_2}V^{\text{eff}}\ket{Q+1,Q+1, m_3m_4} \\
        & \equiv \bra{Q,Q+1,m_1m_2}V\ket{Q,Q+1,m_3m_4}
    \end{split}
\end{equation}
where $\ket{Q,l,m_1m_2}$ is a two-particle state (with the particles not identical) with particle 1 and 2 in the $Y_{Qlm_1}$ and $Y_{Qlm_2}$ orbitals respectively and $V$ is the interaction of the original problem. This definition of $V^{\text{eff}}$ is equivalent to requiring all the Haldane pseudo-potential coefficients $V^{\text{eff}}$ in the LLL to match the corresponding pseudo-potential coefficients of the actual interaction $V$ in the second LL. Note that for the second LL system, we will take the radius of the sphere to be $R = l_B\sqrt{Q}$ and for the LLL system we will take $R^* = l_B\sqrt{Q+1}$, with $l_B$ being the magnetic length.

In this work, we are interested in modelling the Coulomb interaction in the second LL. In particular, as the trial wave functions we are using (for the effective system) are all expressed in real-space (position) representation we then require a real-space form of the effective interaction. An ideal real-space form of the effective interaction would be such that all its pseudo-potential coefficients matched those of $V^{\text{eff}}$ (where it should be noted that there exist many ideal real-space potentials for a given magnetic flux). For a fixed magnetic flux, one can in principle find an effective real-space potential by expressing the effective interaction as a sum of $2Q + 4$ real-space potentials whose vectors of pseudo-potential coefficients are linearly independent. In practice, this can make evaluating the real-space potential, which must be performed many times when using Monte Carlo methods, computationally expensive. Instead in this work, we use an approximate effective interaction, whose real-space form is simple to evaluate. 

The model interaction we use to approximate the ideal effective interaction of the second LL Coulomb interaction is,
\begin{equation} \label{Eq:Veff}
    V^{\text{eff}}(r) = \frac{a_0}{r} + a_1e^{-\alpha_1 r^2} + a_2r^2e^{-\alpha_2 r^2}
\end{equation}
where $r$ is the chord length between the two particles on the sphere, and $a_i$ and $\alpha_i$ are parameters that must be fit. This has been shown in previous works \cite{park_possibility_1998, Sharma2021} to provide a good approximation to the desired effective interaction. In this work, we allow for the parameters $a_i,\alpha_i$ to vary with the system size, where the parameters are determined by minimising the sum of squared differences between the pseudo-potential of this interaction in the LLL and those of the Coulomb interaction in the second LL for $L=0,1,2,\dots,2Q + 2$ (with $2Q$ being the magnetic flux for the system in the second LL). 

\begin{figure}
    \centering
    \includegraphics{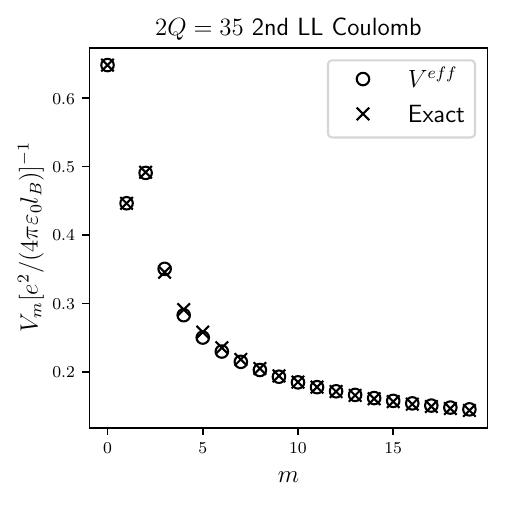}
    \caption{Showing the Haldane pseudo-potential coefficients, $V_m$, of both the Coulomb interaction in the second LL along with the $V_m$ of the fitted LLL $V^{\text{eff}}$, where the total number of magnetic flux quanta is $2Q = 35$ for the system in the second LL. Note that $V_L = V_{m=2Q + 2 - L}$.}
    \label{fig:VeffFitting}
\end{figure}

To fit the effective interaction, we compute the $V_L$'s using Eq. 2.37 of Ref. \cite{Wooten2013}, where the pseudo-potential coefficients are expressed in terms of Wigner 3-$j$ and 6-$j$ symbols, and $V_k$ which are the coefficients in the expansion of $V(r)$ in terms of Legendre polynomials, $V(r) = \sum_k V_k P_k(\cos \theta)$ (with $\theta$ being the angle between the two particles on the sphere). The results of this fitting procedure for the $2Q = 35$ case can be seen in Fig. \ref{fig:VeffFitting}, where it can be seen $V^{\text{eff}}$ can accurately reproduce the Haldane pseudo-potential coefficients for the Coulomb interaction in the 2nd LL with only slight deviations at intermediate $m$. The parameters used for $V^{\text{eff}}$ for the Coulomb interaction in the second LL for the system sizes used in this study are given in Appendix \ref{Sec:VeffParams}. 

\section{Results and discussion} \label{Sec:results}

\subsection{Energetics and overlaps}
\begin{figure*}[t]
    \centering
    \includegraphics{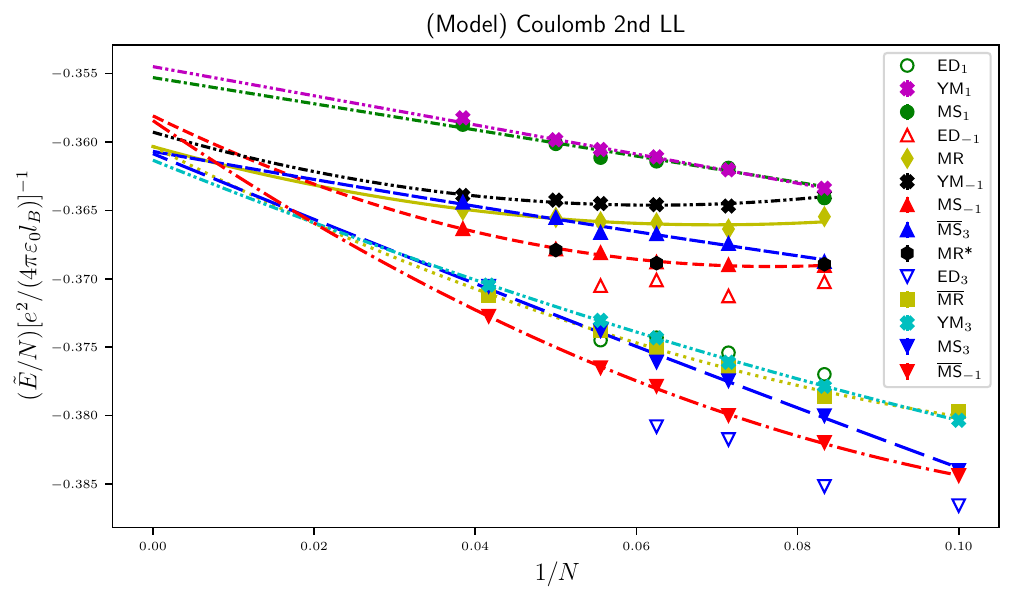}
    \caption{Shows the $\Tilde{E}/N$ (see Eq. \ref{Eq:adjustedE}) for the optimized wave functions as well for the MR, the particle-hole conjugated wave functions of these, denoted with a bar, and the exact ground states $\ED{\ell}$, which are at $L_z = 0$ with total magnetic flux $N_\phi = 2N - 2 + \ell$, for the approximate model (see Sec. \ref{Sec:effInteractions}) of the Coulomb interaction in the 2nd LL. Lines show the resulting polynomial fits in $1/N$ (see main text for full details). Whilst error bars are included for the Monte Carlo estimated energies their sizes are similar to those of the markers.}
    \label{fig:energy}
\end{figure*}

\begin{table}[]
    \centering
    \begin{tabular}{||c|c|c||}
        \hline
        \multirow{2}{*}{state} & \multicolumn{2}{c||}{$(\Tilde{E}/N)[e^2/(4\pi \varepsilon_0 l_B)]^{-1}$} \\
        \cline{2-3}
        & Fit1 & Fit2 \\
        \hline
        MR & $-0.360 \pm 0.001$ & $-0.363 \pm 0.002$ \\ 
$\text{MS}_{-1}$ & $-0.358 \pm 0.002$ & $-0.358 \pm 0.003$ \\ 
$\text{YM}_{-1}$ & $-0.359 \pm 0.001$ & $-0.362 \pm 0.001$ \\ 
$\overline{\text{MS}}_3$ & $-0.361 \pm 0.002$ & $-0.358 \pm 0.003$ \\ 
$\overline{\text{MR}}$ & $-0.360 \pm 0.001$ & $-0.363 \pm 0.002$ \\ 
$\text{MS}_{3}$ & $-0.361 \pm 0.002$ & $-0.358 \pm 0.003$ \\ 
$\text{YM}_3$ & $-0.361 \pm 0.001$ & $-0.362 \pm 0.002$ \\ 
$\overline{\text{MS}}_{-1}$ & $-0.358 \pm 0.002$ & $-0.358 \pm 0.002$ \\ 
$\text{MS}_{1}$ & $-0.355 \pm 0.000$ & $-0.355 \pm 0.000$ \\ 
$\text{YM}_1$ & $-0.354 \pm 0.000$ & $-0.355 \pm 0.000$ \\
        \hline
    \end{tabular}
    \caption{Showing the thermodynamic energies per particle $\Tilde{E}/N$ for the approximate model of the Coulomb interaction in the second LL (see Sec. \ref{Sec:effInteractions}), which are estimated by fitting polynomials to the adjusted energy per particle as a function of $1/N$ (see Fig. \ref{fig:energy}), for the various wave functions. Fit1 uses all data points to fit the given polynomial and Fit2 includes all data points except the smallest system size to fit the given polynomial. Errors have been estimated using the square root of the corresponding diagonal element of the estimated covariance matrix of the polynomial parameters that is outputted by the scipy curve fitting function. Note that the actual errors may be larger as can be seen in the difference between Fit1 and Fit2.}
    \label{tab:energies}
\end{table}

Throughout this section, we will use the shorthand notation $\ED{\ell}$ to denote the exact ground state of the approximate model of the Coulomb interaction in the second Landau level (see Sec. \ref{Sec:effInteractions}) at $L_z = 0$ and for total magnetic flux $N_\phi = 2Q = 2N - 2 + \ell$. Note that at fixed $N$ the $\ED{\ell}$ states for $\ell = -1, 1, 3$ occur at the same total magnetic fluxes as the $\MS{\ell}$ for $\ell = -1, 1, 3$ respectively.

Fig. \ref{fig:energy} shows the \textit{adjusted} energy, $\Tilde{E}$, per particle of the optimized MR* and $\MS{\ell}$ wave functions with $\ell = -1, 1, 3$, as well as the $\Tilde{E}/N$ of the exact ground states, $\ED{\ell}$ for $\ell = -1, 1, 3$, MR, $\YM{\ell}$, with $\ell = -1,1,3$, $\overline{\text{MR}}$ and $\overline{\text{MS}}_{\ell}$, with $\ell = -1, 3$, wave functions as a function of $1/N$ for the approximate model of the Coulomb interaction in the 2nd LL (see Sec. \ref{Sec:effInteractions}), where a bar denotes the particle-hole conjugate of a wave function. The energies of the $\MS{\ell}$ and MR* wave functions have been minimized using the optimization algorithm of Appendix \ref{Sec:optimization}. The adjusted energy is defined by,
\begin{equation} \label{Eq:adjustedE}
    \Tilde{E} = \sqrt{\frac{Q}{N}} \bigg ( E - \frac{N^2e^2}{8\pi \varepsilon_0 l_B \sqrt{Q}} \bigg )
\end{equation}
In the case of the Coulomb interaction, the multiplicative factor adjusts the energies so that the particle density is kept constant, by rescaling the radius of the sphere $R = l_B\sqrt{Q} \rightarrow l_B\sqrt{N}$, and the $\frac{N^2}{8\pi \varepsilon_0 l_B \sqrt{Q}}$ term is the electrostatic energy of a uniformly charged sphere of radius $l_B\sqrt{Q}$ with total charge $Ne$. As discussed in Ref. \cite{morf_monte_1987} this is used to improve the estimates of the thermodynamic energy per particle, which we achieve by fitting a quadratic polynomial in $1/N$ to the $\Tilde{E}/N$ of the MR, $\YM{\ell}$, $\MS{\ell}$ and $\overline{\text{MS}}_{\ell}$, with $\ell = -1, 3$, wave functions and a linear function of $1/N$ for the $\Tilde{E}/N$ of the $\YM{1}$ and $\MS{1}$ wave functions. The resulting estimated thermodynamic energies per particle can be found in Tab. \ref{tab:energies}, where we have included the thermodynamic energy estimates for two different fitting procedures: Fit1 includes all data points to fit the given polynomial and Fit2 which includes all data points except the smallest system size to fit the given polynomial. This is to demonstrate that the thermodynamic energy estimates are somewhat robust. However, it should be noted that the quoted errors in the thermodynamic estimates, given from the output of the curve fitting program, are evidently larger which can be seen from the differences in Fit1 and Fit2. The curves in Fig. \ref{fig:energy} have been fit using the Fit1 procedure. For those wave functions whose real-space form is known exactly, the energies at each system size have been estimated using $\sim 5\times 10^9$ Monte Carlo samples. The energies of the particle-hole conjugated wave functions have been calculated using the result Ref. \cite{moller_composite_2005}, where for a rotationally symmetric interaction under a particle-hole transform the energy transforms as
\begin{equation}
    E \rightarrow \bigg ( 1 - \frac{2N}{2Q + 1} \bigg )E_{\text{filled}} + E
\end{equation}
where $E_{\text{filled}}$ is the energy of the corresponding filled Landau level.

As can be immediately seen from Fig. \ref{fig:energy} despite allowing for some optimization, the $\MS{1}$ wave functions, which are expected to be in the PH-Pfaffian phase, are energetically unfavourable in comparison with the trial wave functions at the other pairing channels tested in this work in the case of the (approximate) Coulomb interaction in the 2nd LL. In fact, the amount by which one can reduce the energy of these $\ell = 1$ wave functions is negligible in comparison with the energy scales in Fig. \ref{fig:energy}.

\begin{table}[]
    \centering
    \begin{tabular}{||c|c|c|c||}
        \hline
         Overlap & $N = 12$ & $N = 14$ & $N = 16$ \\
\hline 
$|\braket{ \text{YM}_{1} | \text{ED}_{1} } |$ & 3(4)\% & 2(4)\% & 2(4)\% \\
$|\braket{ \text{MS}_{1} | \text{ED}_{1} } |$ & 7(4)\% & 2(4)\% & 4(4)\% \\
        \hline
    \end{tabular}
    \caption{Showing overlaps of the $\YM{1}$ and $\MS{1}$ wave functions, expected to be in the PH-Pfaffian phase, with the exact ground state, $\ED{1}$, at the corresponding shift and at $L_z = 0$, in the case of the approximate model interaction for the Coulomb interaction in the first excited Landau level (see Sec. \ref{Sec:effInteractions}).}
    \label{tab:PHOverlaps}
\end{table}

One can also see that the energies of the $\YM{1}$ and $\MS{1}$ wave functions are far higher than the energies of corresponding $\ED{1}$'s. One can also see from Tab. \ref{tab:PHOverlaps} that the overlaps of the $\YM{1}$ and $\MS{1}$ wave functions with the exact ground state for the chosen model interaction, $\ED{1}$, at the corresponding shift, are nearly zero. Thus, the $\YM{1}$ and energy minimized $\MS{1}$ wave functions do not offer a good approximation to the exact ground state.

\begin{table}[h]
    \centering
    \begin{tabular}{||c|c|c|c||}
        \hline
        Overlap & $N = 12$ & $N = 14$ & $N = 16$ \\
\hline 
$|\braket{ \text{MR} | \text{ED}_{-1} } |$ & 76(4)\% & 58(5)\% & 65(4)\% \\
$|\braket{ \text{YM}_{-1} | \text{ED}_{-1} } |$ & 67(4)\% & 45(5)\% & 56(4)\% \\
$|\braket{ \text{MS}_{-1} | \text{ED}_{-1} } |$ & 94(2)\% & 75(6)\% & 88(3)\% \\
        \hline
    \end{tabular}
    \caption{Showing overlaps of the MR, $\YM{-1}$ and the energy minimized $\MS{-1}$ wave functions, expected to be in the Pfaffian phase, with the exact ground state, $\ED{-1}$, at the corresponding shift and at $L_z = 0$, in the case of the approximate model interaction for the Coulomb interaction in the first excited Landau level (see Sec. \ref{Sec:effInteractions}).}
    \label{tab:PfOverlaps}
\end{table}

Another observation that one can make from Fig. \ref{fig:energy} is that the energy minimized $\MS{-1}$ wave functions offer substantial energy reduction over the MR and $\YM{-1}$ wave functions at small system sizes, with energies close to those of the corresponding exact ground states $\ED{-1}$, however still with a notable error. This is further detailed in Tab. \ref{tab:PfOverlaps}, where it can be seen that the energy minimized $\MS{-1}$ wave functions have significantly higher overlaps with the corresponding $\ED{-1}$ states compared with those of the $\YM{-1}$ and MR wave functions, at small system sizes. 

\begin{table}[]
    \centering
    \begin{tabular}{||c|c|c|c||}
         \hline
         Overlap & $N = 10$ & $N = 12$ & $N = 14$ \\
\hline 
$|\braket{ \text{YM}_{3} | \text{ED}_{3} } |$ & 73(4)\% & 52(5)\% & 63(5)\% \\
$|\braket{ \text{MS}_{3} | \text{ED}_{3} } |$ & 89(3)\% & 65(5)\% & 74(4)\% \\
         \hline
    \end{tabular}
    \caption{Showing overlaps of the $\YM{3}$ and the energy minimized $\MS{3}$ wave functions, expected to be in the anti-Pfaffian phase, with the exact ground state, $\ED{3}$, at the corresponding shift and at $L_z = 0$, in the case of the approximate model interaction for the Coulomb interaction in the first excited Landau level (see Sec. \ref{Sec:effInteractions}).}
    \label{tab:aPfOverlaps}
\end{table}

On the other hand, in the $\ell=3$ sector it can also be seen that the energy minimized $\MS{3}$ offer noticeably less energy reduction relative to the other trial states relative to the energy gain of the $\MS{-1}$ wave functions in the $l=1$ sector. The particle-hole conjugated $\MS{-1}$ wave functions, $\overline{\MS{-1}}$, in fact, offer a better approximation to the energies of the corresponding exact ground states $\ED{3}$. As can be seen from Tab. \ref{tab:aPfOverlaps} the $\MS{3}$ wave functions have noticeably better overlaps with $\ED{3}$ compared with $\YM{3}$, although these improvements in the exact ground state overlaps are still not as significant as those of the $\MS{-1}$ wave functions.

Furthermore, as can be seen from Fig. \ref{fig:energy} and Tab. \ref{tab:energies} the energies of the MR, $\overline{\text{MR}}$, $\YM{\ell}$, with $\ell = -1,3$, and the optimized $\MS{\ell}$, along with their particle-hole conjugates $\overline{\text{MS}}_{\ell}$, wave functions all converge in the thermodynamic limit. This is perhaps surprising as one would typically expect that allowing for some energy minimization away from the zero-parameter trial wave functions would allow for an improved estimate of the energy of the actual ground state in the thermodynamic limit. Even the MR* wave functions do not offer a better thermodynamic energy estimate, even though we expect these to always have lower energy than the corresponding MR wave function, which can be seen in Fig. \ref{fig:energy} where they have energies that are not discernable from those of the optimized $\MS{-1}$ wave functions.

There are several possibilities at this point. Firstly, it may, in fact, be the case that the amount by which the energy can be reduced by optimizing the $\MS{-1}$ and $\MS{3}$ wave functions compared with the MR and $\YM{3}$ wave functions falls to zero in the thermodynamic limit. In short, these wave functions may just not offer enough variational freedom at larger system sizes. One may expect, however, that the $\MS{-1}$ wave functions should offer better thermodynamic estimates given that at smaller system sizes their energies are comparable with more exact methods. On the contrary, it should be noted that the polynomial used to extrapolate $\MS{-1}$ has a noticeable curvature, which implies even slight differences between the exact energies and the $\MS{-1}$ can result in a large difference between their thermodynamic extrapolations \footnote{i.e. fitting quadratic polynomials can be much less stable than fitting a linear function}. Of course, one could object to the extrapolation method used here. Whilst it can never be definitively known if the extrapolation method is correct, we can at least verify it is a ``reasonable'' method by the fact that the thermodynamic energy extrapolations of each wave function converge precisely with its corresponding particle-hole conjugate, when the conjugate wave function has been included.

Another possibility is that the optimization algorithm is getting stuck at a local minimum. This could be alleviated by starting the optimization at randomized $g_l$, however, this would come at an increased computational cost as the algorithm must explore a larger area of parameter space to find a minimum. Indeed, the computational cost of optimizing these wave functions using the procedure outlined in Appendix \ref{Sec:optimization} should be emphasised. For example, at $N = 26$ particles the optimization of a wave function with 8 $g_l$ using 300 computing nodes can take around a week, at current computer standards, from the beginning of the optimization to obtaining an accurate estimate of the energy of the optimized wave function. In the worst case, if several fine-tuning phases are required, obtaining the optimum energy at the desired level of accuracy can take on the order of a month. As can be seen from Fig. \ref{fig:energy} this is partly due to the fact that many Monte Carlo samples are required at each iteration of the optimization in order to resolve the rather small differences in the energies of the various wave functions. Thus, although the algorithm could be started from randomized $g_l$ this would come at an increased cost which would render this procedure impractical for most interesting use cases. Whilst, we have checked in Fig. \ref{fig:Optimization} if using extra $g_l$ in the optimizations performed makes very little difference in the energies, it is possible that adding these extra $g_l$ could make a difference to the thermodynamic extrapolation, particularly for the $\MS{-1}$ where the extrapolating polynomial has noticeable curvature. However, with each new $g_l$ the computational cost increases as many more CF orbitals need to be computed. In short, this is perhaps not a practical method for obtaining better thermodynamic estimates in comparison with other numerical methods.

\begin{table}[]
    \centering
    \begin{tabular}{||c|c|c|c||}
        \hline
        Overlap & $N = 10$ & $N = 12$ & $N = 14$ \\
\hline 
$|\braket{ \text{YM}_{3} | \overline{\text{MR}} } |$ & 99(1)\% & 99(1)\% & 98(1)\% \\
        \hline
    \end{tabular}
    \caption{Showing overlaps at different system sizes between the $\YM{3}$ wave function and the particle-hole conjugate of the MR wave function, $\overline{\text{MR}}$, which is a representative wave function of the anti-Pfaffian phase.}
    \label{tab:OverlapMRaPf}
\end{table}

\begin{table}[]
    \centering
    \begin{tabular}{||c|c|c|c||}
        \hline
        Overlap & $N = 12$ & $N = 14$ & $N = 16$ \\
\hline 
$|\braket{ \text{YM}_{-1} | \text{MR} } |$ & 98(1)\% & 98(1)\% & 98(1)\% \\
        \hline
    \end{tabular}
    \caption{Showing overlaps between the $\YM{-1}$ wave function the MR wave function at different system sizes.}
    \label{tab:OverlapMRPf}
\end{table}

Finally, we would also like to emphasise that at small system sizes the $\YM{-1}$ and $\YM{3}$ are very good approximations to the MR and $\overline{\text{MR}}$ wave functions respectively. From Tab. \ref{tab:OverlapMRaPf} one can see that the $\YM{3}$ wave function has an overlap with the $\overline{\text{MR}}$ wave function that is at least 98\% for system sizes of 10-14 particles and from Tab. \ref{tab:OverlapMRPf} that the $\YM{-1}$ wave function has an overlap with the MR wave function of around 98\% for system sizes of 12-16 particles. Such overlaps had already been reported by YM between the $\YM{3}$ and $\overline{\text{MR}}$ wave function, at smaller system sizes, and between the $\YM{-1}$ and MR wave functions at larger system sizes \cite{Yutushui2020}. The fact that these YM wave functions are good approximations to the MR wave function and it's particle-hole conjugate can also be observed from Fig. \ref{fig:energy} where the energies of the $\YM{3}$ and $\YM{-1}$ wave functions are very close to those of the $\overline{\text{MR}}$ and MR wave functions respectively. Whilst the $\YM{-1}$ wave function is clearly not as numerically tractable as the MR wave function, the $\YM{3}$ does provide a more numerically tractable approximation to the $\overline{\text{MR}}$ wave function, at least at small and intermediate system sizes.

\subsection{Effective CF pairing}
\begin{figure*}[ht!]
    \centering
    \includegraphics{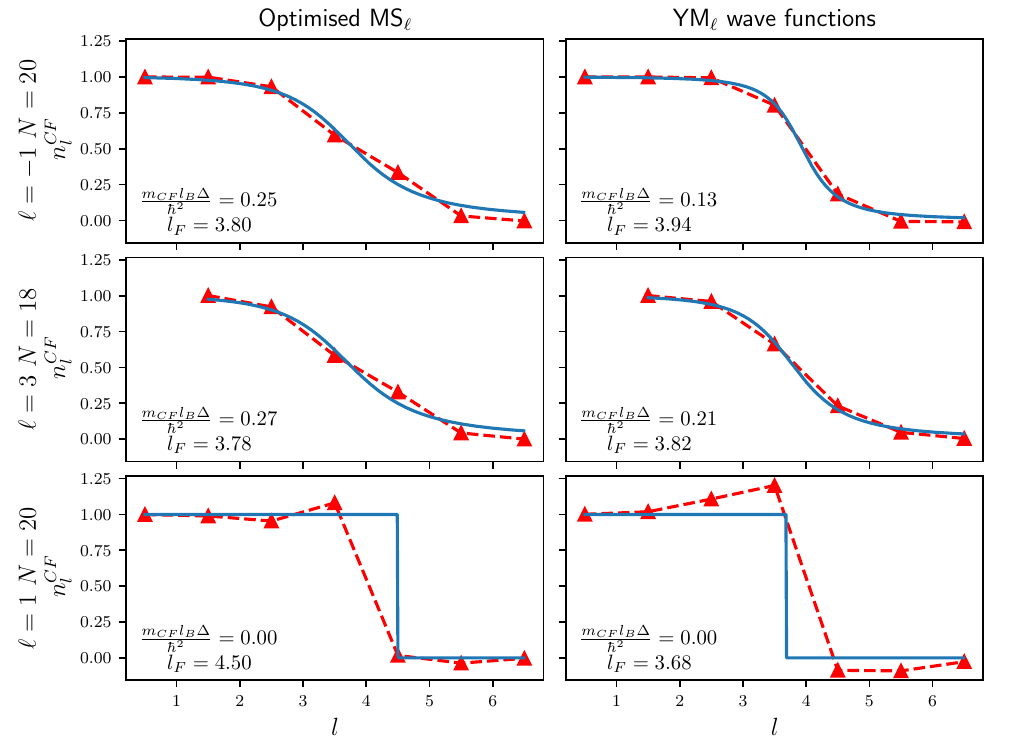}
    \caption{Showing the CF orbital occupation probabilities $n^{\text{CF}}_l$ (see Eq. \ref{Eq:CFn}) for various $\YM{\ell}$ and optimized $\MS{\ell}$ wave functions (dashed lines) along with a corresponding fit to the weak-pairing ansatz of Eq. \ref{Eq:nAnsatz} (solid lines), where $l_F$ is taken to be an adjustable continuous parameter. The fitted parameters are indicated in the bottom left corner of the corresponding plot, where $m_{\text{CF}}$ is the (unknown) effective CF mass. Note that $m_{\text{CF}}l_B\Delta/\hbar^2$ parameter has been estimated where we take the radius of the sphere to be $R = l_B\sqrt{N}$, which keeps the particle density constant.}
    \label{fig:ProbPlot}
\end{figure*}

\begin{figure*}
    \centering
    \includegraphics{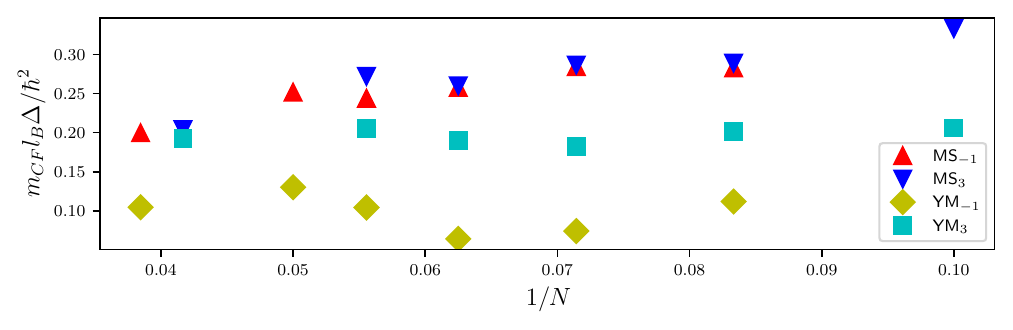}
    \caption{Showing the fitted $m_{\text{CF}}l_B\Delta/\hbar^2$ of the weak pairing ansatz, Eq. \ref{Eq:nAnsatz}, for the effective CF occupations $n_l^{CF}$ for the optimized $\MS{\ell}$ and the fixed parameter $\YM{\ell}$ wave functions for $\ell = -1, 3$ over a variety of system sizes, where $m_{CF}$ is the unknown effective CF mass. Note that $m_{\text{CF}}l_B\Delta/\hbar^2$ parameter has been estimated where we take the radius of the sphere to be $R = l_B\sqrt{N}$, which keeps the particle density constant.}
    \label{fig:deltaPlot}
\end{figure*}

Whilst the optimized $\MS{\ell}$ wave functions do not appear to offer any better thermodynamic energy estimates compared with the $\YM{\ell}$, they do offer more insight at finite-size systems through the effective paired CF description. We have estimated, using Monte Carlo, the effective CF occupation probabilities $n^{\text{CF}}_l$ (Eq. \ref{Eq:CFn}) for the optimized $\MS{\ell}$ wave functions and for the corresponding $\YM{\ell}$. We have then fitted the BCS type weak pairing ansatz (WPA) of Eq. \ref{Eq:nAnsatz} for the various estimated $n^{\text{CF}}_l$, where we allow for the Fermi-level, $l_F$, to be a continuous parameter that can be varied in the fit and we take the radius of the sphere used in Eq. \ref{Eq:nAnsatz} to be $R = l_B\sqrt{N}$ so as to keep the particle density constant. This then allows us to give an estimate for the dimensionless parameter $\frac{m_{\text{CF}}l_B\Delta }{\hbar^2}$, where $m_{\text{CF}}$ is the effective CF mass which we will assume to be constant so that we can take $\frac{m_{\text{CF}}l_B\Delta }{\hbar^2}$ as a measure of the effective CF pairing strength.

Fig. \ref{fig:ProbPlot} shows the estimated $n^{\text{CF}}_l$ and corresponding fitted WPA for the optimized $\MS{\ell}$ and fixed parameter $\YM{\ell}$ wave functions at $\ell = -1, 3, 1$ for $N = 20, 18 , 20$ respectively. The $n^{\text{CF}}_l$ of the $\MS{\ell}$ and $\YM{\ell}$ for $\ell = -1, 3$ can be fit reasonably well to the WPA. 

In Fig. \ref{fig:deltaPlot} shows the fitted $m_{\text{CF}}l_B\Delta/\hbar^2$ parameter of the optimized $\MS{\ell}$ and $\YM{\ell}$ wave functions for $\ell = -1,3$ as a function of $1/N$, where it can generally be seen that the effective CF pairing in the optimized $\MS{\ell}$ wave functions is higher than that of the $\YM{\ell}$. Interestingly we find the effective pairing strength for the optimized $\MS{\ell}$ for $\ell = -1, 3$ wave functions to be roughly the same at the same number of particles $N$, where it should be emphasised that at the same $N$ the $\MS{-1}$ and $\MS{3}$ wave functions occur at different total magnetic flux through the sphere (i.e. have different shifts $\mathcal{S}$).

The data from Fig. \ref{fig:deltaPlot} can also offer more insight into some of the observations one can make from Fig. \ref{fig:energy}. One can see that from Fig. \ref{fig:energy} that the amount by which the energy of the $\MS{-1}$ wave functions can be lowered compared with the corresponding MR wave functions is generally larger than the amount by which the energy of the $\MS{3}$ wave function can be lowered in comparison with the $\YM{3}$ wave functions. Taking the $\YM{-1}$ wave function as an approximation of the MR wave function, this has a simple interpretation in the effective pairing description in that the effective pairing strength of the $\YM{-1}$ wave function is generally lower than that of the $\YM{3}$ wave functions. Thus the $\YM{3}$ wave functions have an effective pairing closer to the optimum compared with the $\YM{-1}$ wave functions which gives some explanation as to why the energy of the $\YM{3}$ wave functions are already close to optimal $\MS{3}$ wave functions, whereas at intermediate system sizes the optimum $\MS{-1}$ wave functions have energy noticeably lower than the corresponding MR wave function. 

Finally, as can also be seen in Fig. \ref{fig:ProbPlot} the $n^{\text{CF}}_l$ of the optimized $\MS{1}$ and $\YM{1}$ wave functions are noticeably larger than one in some cases and are negative in some other cases, which is clearly inconsistent with interpreting these $n^{\text{CF}}_l$ as occupation \textit{probabilites}. This was found to occur at all other tested system sizes. This interpretation of the $n^{\text{CF}}_l$ is based on the assumption that the Jastrow factor of the wave function can be approximated in some ``mean-field'' way, which is usually assumed when interpreting generic CF wave functions. Combining this with the observations of YM, that the $\YM{1}$ wave function shows no sign of emergent pairing in the pair correlation function, and with the evidence that the ``ideal'' PH-Pfaffian wave function may, in fact, represent a gapless phase of matter \cite{pakrouski_approximate_2021}, indicates the possibility that this usual ``mean-field'' interpretation of the Jastrow factor might break down for the current candidate PH-Pfaffian wave functions and so may in fact not be a representative for the phase of matter predicted by Son \cite{Son2015}. It has also recently been argued by Haldane \cite{haldane2023incompressible}, based on a conjecture that FQH states must have a non-zero so-called ``guiding centre quadrupole moment'', that particle-hole symmetric states can never be FQH states, which may give an explanation for these observed pathologies. Despite this, we can still see that the $n^{\text{CF}}_l$ of the optimized $\MS{1}$ wave functions can still be roughly fit to the WPA with $\Delta = 0$, which would correspond to the gapless CF Fermi-liquid. In summary, even after some optimization, we see no evidence of an effective paired CF description for the PH-Pfaffian wave functions.

\section{Conclusion}
In this work, we have shown how the energy of the paired CF wave functions, $\MS{\ell}$, proposed by MS \cite{Moller2008} and extended by YM \cite{Yutushui2020}, can be minimized in a practical manner by varying the pairing function for the pairing channels $\ell = -1, 3, 1$ which are believed to represent the Pfaffian, anti-Pfaffian and PH-Pfaffian topological orders respectively. We have presented the result of such optimizations in the case of an approximate model of the Coulomb interaction in the second Landau level (LL). For pairing channel $\ell = -1$ we found that the energy can be reduced substantially below that of the MR wave function at intermediate system sizes, with a noticeable improvement in the overlaps with the corresponding exact ground state. For the $\ell = 3$ pairing channel, however, we find the energy cannot be reduced as much, although the resulting improvement in the overlaps with the corresponding exact ground state are still notable. We find that optimizing the pairing channel $\ell = 1$ wave functions makes no qualitative difference and these PH-Pfaffian wave functions are still very energetically unfavourable compared with the $\ell = -1, 3$ pairing channels. We have further emphasised that the fixed parameter versions of these wave functions $\YM{\ell}$, proposed by YM \cite{Yutushui2020}, at pairing channels $\ell = -1$ and $\ell = 3$ have very high overlaps (98\%-99\%) with the MR and $\overline{\text{MR}}$ wave functions respectively, at system sizes of around 10-16 particles.

The effective CF pairing of these $\MS{\ell}$ and $\YM{\ell}$ wave functions was then investigated by studying the effective CF orbital occupation probabilities. For the pairing channels $\ell = -1, 3$ we found both the energy minimized $\MS{\ell}$ and $\YM{\ell}$ wave functions show effective CF pairing which can be well approximated by some weak paring BCS type ansatz where the pairing parameters could be extracted, with the $\MS{\ell}$ wave functions showing a stronger pairing strength than the $\YM{\ell}$ wave functions. For the $\YM{1}$ and energy minimized $\MS{1}$ wave functions showed no sign of effective CF pairing, which adds to the growing list of pathologies found for current trial wave functions for the PH-Pfaffian phase. 

It is not clear from this work if the observed convergence of the thermodynamic energies of both the energy minimized and fixed parameter wave functions in the Pfaffian and anti-Pfaffian phases is unique to the Coulomb interaction in the second LL. It would be useful to perform these optimizations at slightly different interactions to see if this convergence is generic. Although one should be cautious of the observed charge density wave phase in the vicinity of the Coulomb interaction in the second LL \cite{rezayi_incompressible_2000}. It would also be interesting to see if minimizing the energy of the $\MS{1}$ wave function for some interaction near the Coulomb interaction will result in an $\ell = 1$ wave function which shows a definite signature of emergent CF pairing.

\begin{acknowledgments}
    Numerical computations have been performed using the DiagHam library. SHS and GJH were partially funded by EPSRC grant EP/S020527/1. GM acknowledges support from the Royal Society under University Research Fellowship URF\textbackslash R\textbackslash 180004. Statement of compliance with EPSRC policy framework on research data: This publication is theoretical work that does not require supporting research data.
\end{acknowledgments}

\appendix

\section{Effective interaction parameters} \label{Sec:VeffParams}
See Table \ref{tab:coulParams} for the $V^{\text{eff}}$ parameters used in this study for the Coulomb interaction in the second LL.

\begin{table}[t]
    \centering
    \begin{tabular}{||c|c|c|c|c|c||}
        \hline
        $2Q$ & $a_0$ & $a_1$ & $\alpha_1$ & $a_2$ & $\alpha_2$ \\
        \hline
        19 & 1.120 & 115.660 & 1.357 & -757.200 & 2.964 \\ 
21 & 1.111 & 140.458 & 1.458 & -973.385 & 3.164 \\ 
23 & 1.102 & 115.604 & 1.353 & -755.689 & 2.959 \\ 
25 & 1.096 & 115.688 & 1.351 & -755.678 & 2.956 \\ 
27 & 1.090 & 115.754 & 1.350 & -755.670 & 2.954 \\ 
29 & 1.085 & 116.224 & 1.350 & -758.751 & 2.954 \\ 
31 & 1.081 & 115.866 & 1.348 & -755.656 & 2.951 \\ 
33 & 1.077 & 115.913 & 1.347 & -755.650 & 2.949 \\ 
35 & 1.073 & 115.957 & 1.346 & -755.645 & 2.948 \\ 
37 & 1.070 & 115.995 & 1.345 & -755.640 & 2.947 \\ 
47 & 1.059 & 116.095 & 1.342 & -755.259 & 2.942 \\ 
49 & 1.057 & 116.178 & 1.342 & -755.617 & 2.941 \\
        \hline
    \end{tabular}
    \caption{Fitted $V^{\text{eff}}$ parameters for the 2nd LL Coulomb interaction at all $2Q$ (of the second LL system) considered in this study.}
    \label{tab:coulParams}
\end{table}

\section{Optimization algorithm} \label{Sec:optimization}
One possible way of minimising the energy of an MS wave function at a given system size would be to use a standard optimization algorithm where the energy for a given set of parameters is computed \textit{exactly}. This is, however, computationally expensive and would restrict one to only working with smaller system sizes. For the energy minimization of wave functions of large systems, there exists a large class of algorithms under the name \textit{variational quantum Monte Carlo} \cite{von_der_linden_quantum_1992}, where Monte Carlo methods are used to estimate the energy and energy gradients. Whilst there are many specific algorithms to choose from, for the problem at hand we have used a combination of the \textit{Stochastic Reconfiguration} (SR) algorithm \cite{Sorella2007} and the Adam optimizer \cite{Kingma2015}, which we will now describe.

First, let us describe the Adam optimizer. Let $E(\mathbf{g})$ be the expectation value of the Energy as a function of the wave function parameters $\mathbf{g}$. We will then impose a yet unspecified geometry on this parameter space defined by some metric tensor $S$ which can vary over the parameter space. At each iteration $t$ of the optimization, we have three vectors, $\mathbf{f}_t \equiv \frac{\partial E}{\partial\mathbf{g}_t}$ (gradient vector), $\mathbf{m}_t$ (momentum) and $\hat{\mathbf{m}}_t$ (bias-corrected momentum). We also have the real numbers $v_t$ and $\hat{v}_t$ the bias-corrected version. The algorithm has four hyper parameters: $\gamma$ (learning rate), $\beta_1$, $\beta_2$ and $\epsilon$. 

Using the momentum $\mathbf{m}_t$ allows the algorithm to smooth out the noise in estimating gradients and $v_t$ ensures we move roughly the same distance in parameter space at every iteration. One can view this algorithm as a particle moving in parameter space with friction and noise in the potential given by $E$. $\beta_1$ can then be thought of as setting our inertia. $\beta_2$ sets the time scale over which we average the sizes of $S^{-1}g_t$. The $\epsilon$ parameter is simply a distance cut-off.

At each iteration, everything is updated by,
\begin{equation}
\begin{split}
    \mathbf{m}_{t + 1} =& \beta_1 \mathbf{m}_{t} + (1-\beta_1)S^{-1}\mathbf{f}_t \\
    \hat{\mathbf{m}}_{t + 1} =& m_{t + 1}/(1 - \beta_1^{t+1}) \\
    v_{t + 1} =& \beta_2 v_t + (1-\beta_2)\mathbf{f}_t^T S^{-1} \mathbf{f}_t \\
    \hat{v}_{t + 1} =& v_{t+1}/(1-\beta_2^{t+1}) \\
    \mathbf{g}_{t + 1} =& \mathbf{g}_t - \gamma \hat{\mathbf{m}}_{t+1}/(\sqrt{\hat{v}_{t+1}} + \epsilon) \\
    \end{split}
\end{equation}
Let $O_l$ be an operator which is diagonal in the position space representation, which is defined by $O_l \equiv \frac{\partial_{g_l}\Psi_{\text{MS}}}{\Psi_{\text{MS}}}$. The energy gradients can then be expressed as,
\begin{equation}
    \frac{\partial E}{\partial g_l} = 2\Re \braket{V^{\text{eff}}O_l} - 2\braket{V^{\text{eff}}}\Re \braket{O_l}
\end{equation}
where $\Re$ denotes the real part. We compute these expectation values using the Hastings-Metropolis Monte Carlo algorithm, which introduces stochastic noise into the optimization.

The SR algorithm can be interpreted as a standard gradient descent algorithm where one uses the Fubini-Study metric to define the geometry of the parameter space of the wave function. To combine this with the Adam optimizer we simply take $S$ to be the Fubini-Study metric. This metric is given by defining the distance between two nearby points in parameter space to be the Hilbert space norm of the difference between the \textit{normalised} states at the two coordinate points. This can be straightforwardly shown to be given by,
\begin{equation}
    S_{l_1l_2} = \Re \braket{O_{l_1}^* O_{l_2}} - \Re\braket{O_{l_1}} \Re \braket{O_{l_2}}
\end{equation}
Using this metric has the advantage that it removes the ambiguity over how the parameters are normalised. At each iteration of the algorithm, we again use the Hastings-Metropolis Monte Carlo algorithm to estimate $S$. 

Both the gradients $\mathbf{f}$ and $S$ are estimated using $\sim 10^{6}$ Monte Carlo samples, with the actual number of samples used increasing with the system size. To lessen the computation time sampling at each iteration was run in parallel using around $10^2$ computing nodes.

In practice, we regularise this metric as the $\Psi_{\text{MS}}$ is invariant under multiplying all the $g_l$ by a constant which will mean $S$ will always have determinant zero. We regularise by, $S_{l_1l_2} \rightarrow ( 1 + \varepsilon \delta_{l_1l_2} )S_{l_1l_2} $, for some small $\varepsilon$ which is kept constant throughout the optimization.

Throughout the optimizations performed in this study we use the recommended hyperparameters from Ref. \cite{Kingma2015} of $\beta_1 = 0.9$, $\beta_2 = 0.999$ and $\epsilon = 10^{-5}$. We found that a suitable learning rate for this optimization problem is around $\gamma \sim 0.005$ and we use the same $\varepsilon$ to regulate $S$ as given in Ref. \cite{Carleo2017} with $\varepsilon = 10^{-3}$. 

\begin{figure}
    \centering
    \includegraphics{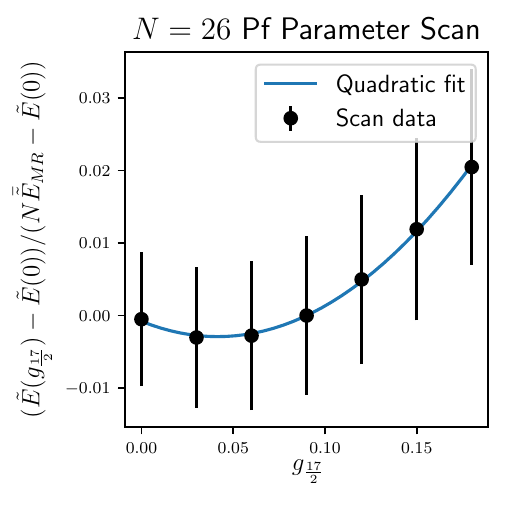}
    \caption{Adjusted energy $\Tilde{E}$ (see Eq. \ref{Eq:adjustedE}) as a function of the 9th $g_l$, around the optimum solution for the first 8 $g_l$ for the $N = 26$ $\MS{-1}$ wave function with the approximate model of the 2nd LL Coulomb interaction, where $\bar{\Tilde{E}}_{MR}$ is the thermodynamic $\Tilde{E}/N$ of the MR wave function (see Tab. \ref{tab:energies} where we use the Fit1 value).}
    \label{fig:Optimization}
\end{figure}

\begin{figure}
    \centering
    \includegraphics{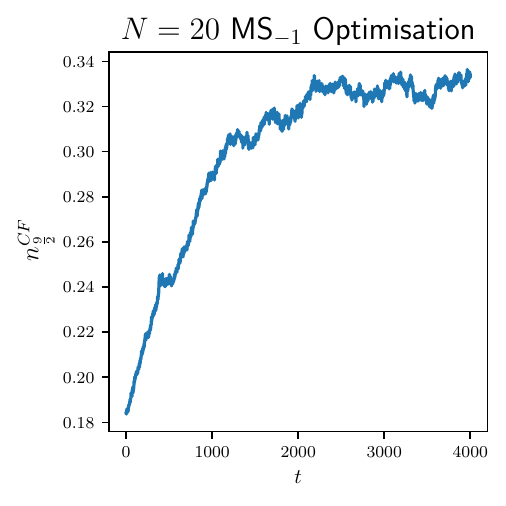}
    \caption{Showing $n^{\text{CF}}_{\frac{9}{2}}$ (see Eq. \ref{Eq:CFn}) at each iteration $t$ of the optimization of the $N = 20$ $\MS{-1}$ wave function with the approximate model of the 2nd LL Coulomb interaction.}
    \label{fig:Optimization2}
\end{figure}

The issue of how to pick the number of $g_l$ to use is addressed as follows. For the $\MS{-1}$ case, we ran the optimization at $N=12$ with the first $7$ $g_l$, where it was found that the broadening of the Fermi surface, as seen through the $n^{\text{CF}}_l$, was around 2 in $l$ space (i.e. $n^\text{CF}_l$ were found to be sufficiently close to zero for $l > l_f + 2$). We expect this broadening in $l$ space to scale as $\sqrt{N}$ as $l \sim kR$. Thus, for system sizes from 12-20 particles, we used the first 7 $g_l$ to optimize the $\MS{-1}$ wave functions and at 26 particles we used the first 8. To demonstrate this number of parameters is sufficient we estimated the energy as a function of the $9^{\text{th}}$ $g_l$ around the optimum set when using the first 8 $g_l$ for the $N=26$ $\MS{-1}$ optimization with the approximate model of the Coulomb interaction in the second LL. The result of this parameter scan can be seen in the plot of Fig. \ref{fig:Optimization}, where it can be seen that the possible energy reduction per particle that can be achieved by varying this parameter is small in comparison to the difference between the previous optimum energy per particle and the thermodynamic energy per particle of the Moore-Read wave function (see Fig. \ref{fig:energy} and Tab. \ref{tab:energies} for details of thermodynamic estimates where we use the Fit1 estimate for the thermodynamic energy per particle of the MR wave function). Similarly, for both the MR* and the $\MS{1}$ wave functions we used the first 7 $g_l$ for 12-20 particles and used the first 8 for 26 particles. By the same method, we found that using the first 6 and 7 $g_l$ was sufficient to optimize the $\MS{3}$ wave functions for 10-18 and 24 particles respectively. 

When optimizing a given $\MS{\ell}$ wave function we start the algorithm with $g_l = \frac{1}{2l + 1}$ for the parameters that are actually varied and all other $g_l$ are set to zero, $g_l = 0$. This is the approximate $\YM{\ell}$ wave function which we expect to be close to the minimum energy.

The plot of Fig. \ref{fig:Optimization2} shows the $n^{\text{CF}}_{\frac{9}{2}}$ occupation probability at each iteration $t$ for the optimization of the $N = 20$ $\MS{-1}$ wave function with the approximate model of the 2nd LL Coulomb interaction (note that $l = \frac{9}{2}$ is just at the Fermi level). Initially, the algorithm moves sharply towards a minimum where it plateaus near the minimum energy at around $t = 4000$. We found that this algorithm converged after around $t \sim 4000$ iterations for all cases considered using the hyperparameters given above. One can also see that the apparent noise in the $n^{\text{CF}}_{\frac{9}{2}}(t)$ path increases as we approach the minimum. At larger system sizes this sometimes required a fine-tuning stage with around 100 iterations where the number of samples and the learning rate are increased to obtain the minimum energy solution with the desired accuracy.

\bibliography{Refs.bib}

\begin{thebibliography}{51}%
\makeatletter
\providecommand \@ifxundefined [1]{%
 \@ifx{#1\undefined}
}%
\providecommand \@ifnum [1]{%
 \ifnum #1\expandafter \@firstoftwo
 \else \expandafter \@secondoftwo
 \fi
}%
\providecommand \@ifx [1]{%
 \ifx #1\expandafter \@firstoftwo
 \else \expandafter \@secondoftwo
 \fi
}%
\providecommand \natexlab [1]{#1}%
\providecommand \enquote  [1]{``#1''}%
\providecommand \bibnamefont  [1]{#1}%
\providecommand \bibfnamefont [1]{#1}%
\providecommand \citenamefont [1]{#1}%
\providecommand \href@noop [0]{\@secondoftwo}%
\providecommand \href [0]{\begingroup \@sanitize@url \@href}%
\providecommand \@href[1]{\@@startlink{#1}\@@href}%
\providecommand \@@href[1]{\endgroup#1\@@endlink}%
\providecommand \@sanitize@url [0]{\catcode `\\12\catcode `\$12\catcode
  `\&12\catcode `\#12\catcode `\^12\catcode `\_12\catcode `\%12\relax}%
\providecommand \@@startlink[1]{}%
\providecommand \@@endlink[0]{}%
\providecommand \url  [0]{\begingroup\@sanitize@url \@url }%
\providecommand \@url [1]{\endgroup\@href {#1}{\urlprefix }}%
\providecommand \urlprefix  [0]{URL }%
\providecommand \Eprint [0]{\href }%
\providecommand \doibase [0]{https://doi.org/}%
\providecommand \selectlanguage [0]{\@gobble}%
\providecommand \bibinfo  [0]{\@secondoftwo}%
\providecommand \bibfield  [0]{\@secondoftwo}%
\providecommand \translation [1]{[#1]}%
\providecommand \BibitemOpen [0]{}%
\providecommand \bibitemStop [0]{}%
\providecommand \bibitemNoStop [0]{.\EOS\space}%
\providecommand \EOS [0]{\spacefactor3000\relax}%
\providecommand \BibitemShut  [1]{\csname bibitem#1\endcsname}%
\let\auto@bib@innerbib\@empty
\bibitem [{\citenamefont {Willett}\ \emph {et~al.}(1987)\citenamefont
  {Willett}, \citenamefont {Eisenstein}, \citenamefont {St{\"o}rmer},
  \citenamefont {Tsui}, \citenamefont {Gossard},\ and\ \citenamefont
  {English}}]{willett_observation_1987}%
  \BibitemOpen
  \bibfield  {author} {\bibinfo {author} {\bibfnamefont {R.}~\bibnamefont
  {Willett}}, \bibinfo {author} {\bibfnamefont {J.~P.}\ \bibnamefont
  {Eisenstein}}, \bibinfo {author} {\bibfnamefont {H.~L.}\ \bibnamefont
  {St{\"o}rmer}}, \bibinfo {author} {\bibfnamefont {D.~C.}\ \bibnamefont
  {Tsui}}, \bibinfo {author} {\bibfnamefont {A.~C.}\ \bibnamefont {Gossard}},\
  and\ \bibinfo {author} {\bibfnamefont {J.~H.}\ \bibnamefont {English}},\
  }\bibfield  {title} {\bibinfo {title} {Observation of an even-denominator
  quantum number in the fractional quantum {Hall} effect},\ }\href
  {https://doi.org/10.1103/PhysRevLett.59.1776} {\bibfield  {journal} {\bibinfo
   {journal} {Phys. Rev. Lett.}\ }\textbf {\bibinfo {volume} {59}},\ \bibinfo
  {pages} {1776} (\bibinfo {year} {1987})}\BibitemShut {NoStop}%
\bibitem [{\citenamefont {Moore}\ and\ \citenamefont {Read}(1991)}]{Moore1991}%
  \BibitemOpen
  \bibfield  {author} {\bibinfo {author} {\bibfnamefont {G.}~\bibnamefont
  {Moore}}\ and\ \bibinfo {author} {\bibfnamefont {N.}~\bibnamefont {Read}},\
  }\bibfield  {title} {\bibinfo {title} {Nonabelions in the fractional quantum
  hall effect},\ }\href {https://doi.org/10.1016/0550-3213(91)90407-O}
  {\bibfield  {journal} {\bibinfo  {journal} {Nuc. Phys. B}\ }\textbf {\bibinfo
  {volume} {360}},\ \bibinfo {pages} {362} (\bibinfo {year}
  {1991})}\BibitemShut {NoStop}%
\bibitem [{\citenamefont {Storni}\ \emph {et~al.}(2010)\citenamefont {Storni},
  \citenamefont {Morf},\ and\ \citenamefont
  {Das~Sarma}}]{storni_fractional_2010}%
  \BibitemOpen
  \bibfield  {author} {\bibinfo {author} {\bibfnamefont {M.}~\bibnamefont
  {Storni}}, \bibinfo {author} {\bibfnamefont {R.~H.}\ \bibnamefont {Morf}},\
  and\ \bibinfo {author} {\bibfnamefont {S.}~\bibnamefont {Das~Sarma}},\
  }\bibfield  {title} {\bibinfo {title} {Fractional {Quantum} {Hall} {State} at
  $\nu$ = 5/2 and the {Moore}-{Read} {Pfaffian}},\ }\href
  {https://link.aps.org/doi/10.1103/PhysRevLett.104.076803} {\bibfield
  {journal} {\bibinfo  {journal} {Phys. Rev. Lett.}\ }\textbf {\bibinfo
  {volume} {104}},\ \bibinfo {pages} {076803} (\bibinfo {year}
  {2010})}\BibitemShut {NoStop}%
\bibitem [{\citenamefont {Levin}\ \emph {et~al.}(2007)\citenamefont {Levin},
  \citenamefont {Halperin},\ and\ \citenamefont
  {Rosenow}}]{levin_particle-hole_2007}%
  \BibitemOpen
  \bibfield  {author} {\bibinfo {author} {\bibfnamefont {M.}~\bibnamefont
  {Levin}}, \bibinfo {author} {\bibfnamefont {B.~I.}\ \bibnamefont
  {Halperin}},\ and\ \bibinfo {author} {\bibfnamefont {B.}~\bibnamefont
  {Rosenow}},\ }\bibfield  {title} {\bibinfo {title} {Particle-{Hole}
  {Symmetry} and the {Pfaffian} {State}},\ }\href
  {https://link.aps.org/doi/10.1103/PhysRevLett.99.236806} {\bibfield
  {journal} {\bibinfo  {journal} {Phys. Rev. Lett.}\ }\textbf {\bibinfo
  {volume} {99}},\ \bibinfo {pages} {236806} (\bibinfo {year}
  {2007})}\BibitemShut {NoStop}%
\bibitem [{\citenamefont {Lee}\ \emph {et~al.}(2007)\citenamefont {Lee},
  \citenamefont {Ryu}, \citenamefont {Nayak},\ and\ \citenamefont
  {Fisher}}]{lee_particle-hole_2007}%
  \BibitemOpen
  \bibfield  {author} {\bibinfo {author} {\bibfnamefont {S.-S.}\ \bibnamefont
  {Lee}}, \bibinfo {author} {\bibfnamefont {S.}~\bibnamefont {Ryu}}, \bibinfo
  {author} {\bibfnamefont {C.}~\bibnamefont {Nayak}},\ and\ \bibinfo {author}
  {\bibfnamefont {M.~P.~A.}\ \bibnamefont {Fisher}},\ }\bibfield  {title}
  {\bibinfo {title} {Particle-{Hole} {Symmetry} and the $\nu$ = 5/2 {Quantum}
  {Hall} {State}},\ }\href
  {https://link.aps.org/doi/10.1103/PhysRevLett.99.236807} {\bibfield
  {journal} {\bibinfo  {journal} {Phys. Rev. Lett.}\ }\textbf {\bibinfo
  {volume} {99}},\ \bibinfo {pages} {236807} (\bibinfo {year}
  {2007})}\BibitemShut {NoStop}%
\bibitem [{\citenamefont {Son}(2015)}]{Son2015}%
  \BibitemOpen
  \bibfield  {author} {\bibinfo {author} {\bibfnamefont {D.~T.}\ \bibnamefont
  {Son}},\ }\bibfield  {title} {\bibinfo {title} {Is the composite fermion a
  dirac particle?},\ }\href@noop {} {\bibfield  {journal} {\bibinfo  {journal}
  {Phys. Rev. X}\ }\textbf {\bibinfo {volume} {5}} (\bibinfo {year}
  {2015})}\BibitemShut {NoStop}%
\bibitem [{\citenamefont {Greiter}\ \emph {et~al.}(1991)\citenamefont
  {Greiter}, \citenamefont {Wen},\ and\ \citenamefont
  {Wilczek}}]{greiter_paired_1991}%
  \BibitemOpen
  \bibfield  {author} {\bibinfo {author} {\bibfnamefont {M.}~\bibnamefont
  {Greiter}}, \bibinfo {author} {\bibfnamefont {X.~G.}\ \bibnamefont {Wen}},\
  and\ \bibinfo {author} {\bibfnamefont {F.}~\bibnamefont {Wilczek}},\
  }\bibfield  {title} {\bibinfo {title} {Paired {Hall} state at half filling},\
  }\href {https://link.aps.org/doi/10.1103/PhysRevLett.66.3205} {\bibfield
  {journal} {\bibinfo  {journal} {Phys. Rev. Lett.}\ }\textbf {\bibinfo
  {volume} {66}},\ \bibinfo {pages} {3205} (\bibinfo {year}
  {1991})}\BibitemShut {NoStop}%
\bibitem [{\citenamefont {Park}\ \emph {et~al.}(1998)\citenamefont {Park},
  \citenamefont {Melik-Alaverdian}, \citenamefont {Bonesteel},\ and\
  \citenamefont {Jain}}]{park_possibility_1998}%
  \BibitemOpen
  \bibfield  {author} {\bibinfo {author} {\bibfnamefont {K.}~\bibnamefont
  {Park}}, \bibinfo {author} {\bibfnamefont {V.}~\bibnamefont
  {Melik-Alaverdian}}, \bibinfo {author} {\bibfnamefont {N.~E.}\ \bibnamefont
  {Bonesteel}},\ and\ \bibinfo {author} {\bibfnamefont {J.~K.}\ \bibnamefont
  {Jain}},\ }\bibfield  {title} {\bibinfo {title} {Possibility of p-wave
  pairing of composite fermions at 1 2},\ }\href
  {https://journals.aps.org/prb/abstract/10.1103/PhysRevB.58.R10167} {\bibfield
   {journal} {\bibinfo  {journal} {Phys. Rev. B}\ }\textbf {\bibinfo {volume}
  {58}} (\bibinfo {year} {1998})}\BibitemShut {NoStop}%
\bibitem [{\citenamefont {Read}\ and\ \citenamefont {Green}(2000)}]{Read2000}%
  \BibitemOpen
  \bibfield  {author} {\bibinfo {author} {\bibfnamefont {N.}~\bibnamefont
  {Read}}\ and\ \bibinfo {author} {\bibfnamefont {D.}~\bibnamefont {Green}},\
  }\bibfield  {title} {\bibinfo {title} {Paired states of fermions in two
  dimensions with breaking of parity and time-reversal symmetries and the
  fractional quantum {Hall} effect},\ }\href@noop {} {\bibfield  {journal}
  {\bibinfo  {journal} {Phys. Rev. B}\ }\textbf {\bibinfo {volume} {61}},\
  \bibinfo {pages} {10267} (\bibinfo {year} {2000})}\BibitemShut {NoStop}%
\bibitem [{\citenamefont {Barkeshli}\ \emph {et~al.}(2015)\citenamefont
  {Barkeshli}, \citenamefont {Mulligan},\ and\ \citenamefont
  {Fisher}}]{barkeshli2015particle}%
  \BibitemOpen
  \bibfield  {author} {\bibinfo {author} {\bibfnamefont {M.}~\bibnamefont
  {Barkeshli}}, \bibinfo {author} {\bibfnamefont {M.}~\bibnamefont
  {Mulligan}},\ and\ \bibinfo {author} {\bibfnamefont {M.~P.}\ \bibnamefont
  {Fisher}},\ }\bibfield  {title} {\bibinfo {title} {Particle-hole symmetry and
  the composite fermi liquid},\ }\href
  {https://journals.aps.org/prb/abstract/10.1103/PhysRevB.92.165125} {\bibfield
   {journal} {\bibinfo  {journal} {Phys. Rev. B}\ }\textbf {\bibinfo {volume}
  {92}},\ \bibinfo {pages} {165125} (\bibinfo {year} {2015})}\BibitemShut
  {NoStop}%
\bibitem [{\citenamefont {Zucker}\ and\ \citenamefont
  {Feldman}(2016)}]{zucker_stabilization_2016}%
  \BibitemOpen
  \bibfield  {author} {\bibinfo {author} {\bibfnamefont {P.~T.}\ \bibnamefont
  {Zucker}}\ and\ \bibinfo {author} {\bibfnamefont {D.~E.}\ \bibnamefont
  {Feldman}},\ }\bibfield  {title} {\bibinfo {title} {Stabilization of the
  {Particle}-{Hole} {Pfaffian} {Order} by {Landau}-{Level} {Mixing} and
  {Impurities} {That} {Break} {Particle}-{Hole} {Symmetry}},\ }\href
  {https://link.aps.org/doi/10.1103/PhysRevLett.117.096802} {\bibfield
  {journal} {\bibinfo  {journal} {Phys. Rev. Lett.}\ }\textbf {\bibinfo
  {volume} {117}},\ \bibinfo {pages} {096802} (\bibinfo {year}
  {2016})}\BibitemShut {NoStop}%
\bibitem [{\citenamefont {Morf}(1998)}]{morf_transition_1998}%
  \BibitemOpen
  \bibfield  {author} {\bibinfo {author} {\bibfnamefont {R.~H.}\ \bibnamefont
  {Morf}},\ }\bibfield  {title} {\bibinfo {title} {Transition from quantum hall
  to compressible states in the second landau level: new light on the $\nu$=
  5/2 enigma},\ }\href@noop {} {\bibfield  {journal} {\bibinfo  {journal}
  {Phys. Rev. Lett.}\ }\textbf {\bibinfo {volume} {80}},\ \bibinfo {pages}
  {1505} (\bibinfo {year} {1998})}\BibitemShut {NoStop}%
\bibitem [{\citenamefont {Rezayi}\ and\ \citenamefont
  {Haldane}(2000)}]{rezayi_incompressible_2000}%
  \BibitemOpen
  \bibfield  {author} {\bibinfo {author} {\bibfnamefont {E.~H.}\ \bibnamefont
  {Rezayi}}\ and\ \bibinfo {author} {\bibfnamefont {F.~D.~M.}\ \bibnamefont
  {Haldane}},\ }\bibfield  {title} {\bibinfo {title} {Incompressible {Paired}
  {Hall} {State}, {Stripe} {Order}, and the {Composite} {Fermion} {Liquid}
  {Phase} in {Half}-{Filled} {Landau} {Levels}},\ }\href
  {https://link.aps.org/doi/10.1103/PhysRevLett.84.4685} {\bibfield  {journal}
  {\bibinfo  {journal} {Phys. Rev. Lett.}\ }\textbf {\bibinfo {volume} {84}},\
  \bibinfo {pages} {4685} (\bibinfo {year} {2000})}\BibitemShut {NoStop}%
\bibitem [{\citenamefont {Feiguin}\ \emph {et~al.}(2008)\citenamefont
  {Feiguin}, \citenamefont {Rezayi}, \citenamefont {Nayak},\ and\ \citenamefont
  {Das~Sarma}}]{feiguin_density_2008}%
  \BibitemOpen
  \bibfield  {author} {\bibinfo {author} {\bibfnamefont {A.~E.}\ \bibnamefont
  {Feiguin}}, \bibinfo {author} {\bibfnamefont {E.}~\bibnamefont {Rezayi}},
  \bibinfo {author} {\bibfnamefont {C.}~\bibnamefont {Nayak}},\ and\ \bibinfo
  {author} {\bibfnamefont {S.}~\bibnamefont {Das~Sarma}},\ }\bibfield  {title}
  {\bibinfo {title} {Density {Matrix} {Renormalization} {Group} {Study} of
  {Incompressible} {Fractional} {Quantum} {Hall} {States}},\ }\href
  {https://link.aps.org/doi/10.1103/PhysRevLett.100.166803} {\bibfield
  {journal} {\bibinfo  {journal} {Phys. Rev. Lett.}\ }\textbf {\bibinfo
  {volume} {100}},\ \bibinfo {pages} {166803} (\bibinfo {year}
  {2008})}\BibitemShut {NoStop}%
\bibitem [{\citenamefont {Rezayi}\ and\ \citenamefont
  {Simon}(2011)}]{Rezayi2011}%
  \BibitemOpen
  \bibfield  {author} {\bibinfo {author} {\bibfnamefont {E.~H.}\ \bibnamefont
  {Rezayi}}\ and\ \bibinfo {author} {\bibfnamefont {S.~H.}\ \bibnamefont
  {Simon}},\ }\bibfield  {title} {\bibinfo {title} {Breaking of particle-hole
  symmetry by landau level mixing in the $\nu$=5/2 quantized hall state},\
  }\href {https://doi.org/10.1103/PhysRevLett.106.116801} {\bibfield  {journal}
  {\bibinfo  {journal} {Phys. Rev. Lett.}\ }\textbf {\bibinfo {volume} {106}},\
  \bibinfo {pages} {116801} (\bibinfo {year} {2011})}\BibitemShut {NoStop}%
\bibitem [{\citenamefont {Simon}\ and\ \citenamefont
  {Rezayi}(2013)}]{simon_landau_2013}%
  \BibitemOpen
  \bibfield  {author} {\bibinfo {author} {\bibfnamefont {S.~H.}\ \bibnamefont
  {Simon}}\ and\ \bibinfo {author} {\bibfnamefont {E.~H.}\ \bibnamefont
  {Rezayi}},\ }\bibfield  {title} {\bibinfo {title} {Landau level mixing in the
  perturbative limit},\ }\href
  {https://link.aps.org/doi/10.1103/PhysRevB.87.155426} {\bibfield  {journal}
  {\bibinfo  {journal} {Phys. Rev. B}\ }\textbf {\bibinfo {volume} {87}},\
  \bibinfo {pages} {155426} (\bibinfo {year} {2013})}\BibitemShut {NoStop}%
\bibitem [{\citenamefont {Zaletel}\ \emph {et~al.}(2015)\citenamefont
  {Zaletel}, \citenamefont {Mong}, \citenamefont {Pollmann},\ and\
  \citenamefont {Rezayi}}]{Zaletel2015}%
  \BibitemOpen
  \bibfield  {author} {\bibinfo {author} {\bibfnamefont {M.~P.}\ \bibnamefont
  {Zaletel}}, \bibinfo {author} {\bibfnamefont {R.~S.}\ \bibnamefont {Mong}},
  \bibinfo {author} {\bibfnamefont {F.}~\bibnamefont {Pollmann}},\ and\
  \bibinfo {author} {\bibfnamefont {E.~H.}\ \bibnamefont {Rezayi}},\ }\bibfield
   {title} {\bibinfo {title} {Infinite density matrix renormalization group for
  multicomponent quantum {Hall} systems},\ }\href
  {https://doi.org/10.1103/PhysRevB.91.045115} {\bibfield  {journal} {\bibinfo
  {journal} {Phys. Rev. B}\ }\textbf {\bibinfo {volume} {91}},\ \bibinfo
  {pages} {045115} (\bibinfo {year} {2015})},\ \bibinfo {note} {arXiv:
  1410.3861 Publisher: American Physical Society}\BibitemShut {NoStop}%
\bibitem [{\citenamefont {Rezayi}(2017)}]{rezayi_landau_2017}%
  \BibitemOpen
  \bibfield  {author} {\bibinfo {author} {\bibfnamefont {E.~H.}\ \bibnamefont
  {Rezayi}},\ }\bibfield  {title} {\bibinfo {title} {Landau {Level} {Mixing}
  and the {Ground} {State} of the $\nu=5/2$ {Quantum} {Hall} {Effect}},\ }\href
  {https://link.aps.org/doi/10.1103/PhysRevLett.119.026801} {\bibfield
  {journal} {\bibinfo  {journal} {Phys. Rev. Lett.}\ }\textbf {\bibinfo
  {volume} {119}},\ \bibinfo {pages} {026801} (\bibinfo {year}
  {2017})}\BibitemShut {NoStop}%
\bibitem [{\citenamefont {Rezayi}\ \emph {et~al.}(2021)\citenamefont {Rezayi},
  \citenamefont {Pakrouski},\ and\ \citenamefont {Haldane}}]{Rezayi2021}%
  \BibitemOpen
  \bibfield  {author} {\bibinfo {author} {\bibfnamefont {E.~H.}\ \bibnamefont
  {Rezayi}}, \bibinfo {author} {\bibfnamefont {K.}~\bibnamefont {Pakrouski}},\
  and\ \bibinfo {author} {\bibfnamefont {F.~D.~M.}\ \bibnamefont {Haldane}},\
  }\bibfield  {title} {\bibinfo {title} {Stability of the particle-hole
  pfaffian state and the $\frac{5}{2}$-fractional quantum hall effect},\ }\href
  {https://doi.org/10.1103/physrevb.104.l081407} {\bibfield  {journal}
  {\bibinfo  {journal} {Phys. Rev. B}\ }\textbf {\bibinfo {volume} {104}},\
  \bibinfo {pages} {81407} (\bibinfo {year} {2021})}\BibitemShut {NoStop}%
\bibitem [{\citenamefont {Banerjee}\ \emph {et~al.}(2018)\citenamefont
  {Banerjee}, \citenamefont {Heiblum}, \citenamefont {Umansky}, \citenamefont
  {Feldman}, \citenamefont {Oreg},\ and\ \citenamefont
  {Stern}}]{banerjee_observation_2018}%
  \BibitemOpen
  \bibfield  {author} {\bibinfo {author} {\bibfnamefont {M.}~\bibnamefont
  {Banerjee}}, \bibinfo {author} {\bibfnamefont {M.}~\bibnamefont {Heiblum}},
  \bibinfo {author} {\bibfnamefont {V.}~\bibnamefont {Umansky}}, \bibinfo
  {author} {\bibfnamefont {D.~E.}\ \bibnamefont {Feldman}}, \bibinfo {author}
  {\bibfnamefont {Y.}~\bibnamefont {Oreg}},\ and\ \bibinfo {author}
  {\bibfnamefont {A.}~\bibnamefont {Stern}},\ }\bibfield  {title} {\bibinfo
  {title} {Observation of half-integer thermal {Hall} conductance},\ }\href
  {https://www.nature.com/articles/s41586-018-0184-1} {\bibfield  {journal}
  {\bibinfo  {journal} {Nature}\ }\textbf {\bibinfo {volume} {559}},\ \bibinfo
  {pages} {205} (\bibinfo {year} {2018})}\BibitemShut {NoStop}%
\bibitem [{\citenamefont {Dutta}\ \emph {et~al.}(2021)\citenamefont {Dutta},
  \citenamefont {Yang}, \citenamefont {Melcer}, \citenamefont {Kundu},
  \citenamefont {Heiblum}, \citenamefont {Umansky}, \citenamefont {Oreg},
  \citenamefont {Stern},\ and\ \citenamefont {Mross}}]{Dutta2021}%
  \BibitemOpen
  \bibfield  {author} {\bibinfo {author} {\bibfnamefont {B.}~\bibnamefont
  {Dutta}}, \bibinfo {author} {\bibfnamefont {W.}~\bibnamefont {Yang}},
  \bibinfo {author} {\bibfnamefont {R.~A.}\ \bibnamefont {Melcer}}, \bibinfo
  {author} {\bibfnamefont {H.~K.}\ \bibnamefont {Kundu}}, \bibinfo {author}
  {\bibfnamefont {M.}~\bibnamefont {Heiblum}}, \bibinfo {author} {\bibfnamefont
  {V.}~\bibnamefont {Umansky}}, \bibinfo {author} {\bibfnamefont
  {Y.}~\bibnamefont {Oreg}}, \bibinfo {author} {\bibfnamefont {A.}~\bibnamefont
  {Stern}},\ and\ \bibinfo {author} {\bibfnamefont {D.}~\bibnamefont {Mross}},\
  }\bibfield  {title} {\bibinfo {title} {Novel method distinguishing between
  competing topological orders},\ }\href {https://arxiv.org/abs/2101.01419v1}
  {\bibfield  {journal} {\bibinfo  {journal} {arXiv preprint arXiv:2101.01419}\
  } (\bibinfo {year} {2021})}\BibitemShut {NoStop}%
\bibitem [{\citenamefont {Mross}\ \emph {et~al.}(2018)\citenamefont {Mross},
  \citenamefont {Oreg}, \citenamefont {Stern}, \citenamefont {Margalit},\ and\
  \citenamefont {Heiblum}}]{mross_theory_2018}%
  \BibitemOpen
  \bibfield  {author} {\bibinfo {author} {\bibfnamefont {D.~F.}\ \bibnamefont
  {Mross}}, \bibinfo {author} {\bibfnamefont {Y.}~\bibnamefont {Oreg}},
  \bibinfo {author} {\bibfnamefont {A.}~\bibnamefont {Stern}}, \bibinfo
  {author} {\bibfnamefont {G.}~\bibnamefont {Margalit}},\ and\ \bibinfo
  {author} {\bibfnamefont {M.}~\bibnamefont {Heiblum}},\ }\bibfield  {title}
  {\bibinfo {title} {Theory of {Disorder}-{Induced} {Half}-{Integer} {Thermal}
  {Hall} {Conductance}},\ }\href
  {https://link.aps.org/doi/10.1103/PhysRevLett.121.026801} {\bibfield
  {journal} {\bibinfo  {journal} {Phys. Rev. Lett.}\ }\textbf {\bibinfo
  {volume} {121}},\ \bibinfo {pages} {026801} (\bibinfo {year}
  {2018})}\BibitemShut {NoStop}%
\bibitem [{\citenamefont {Lian}\ and\ \citenamefont
  {Wang}(2018)}]{lian_theory_2018}%
  \BibitemOpen
  \bibfield  {author} {\bibinfo {author} {\bibfnamefont {B.}~\bibnamefont
  {Lian}}\ and\ \bibinfo {author} {\bibfnamefont {J.}~\bibnamefont {Wang}},\
  }\bibfield  {title} {\bibinfo {title} {Theory of the disordered $\nu$ = 5/2
  quantum thermal {Hall} state: {Emergent} symmetry and phase diagram},\ }\href
  {https://link.aps.org/doi/10.1103/PhysRevB.97.165124} {\bibfield  {journal}
  {\bibinfo  {journal} {Phys. Rev. B}\ }\textbf {\bibinfo {volume} {97}},\
  \bibinfo {pages} {165124} (\bibinfo {year} {2018})}\BibitemShut {NoStop}%
\bibitem [{\citenamefont {Wang}\ \emph {et~al.}(2018)\citenamefont {Wang},
  \citenamefont {Vishwanath},\ and\ \citenamefont
  {Halperin}}]{wang_topological_2018}%
  \BibitemOpen
  \bibfield  {author} {\bibinfo {author} {\bibfnamefont {C.}~\bibnamefont
  {Wang}}, \bibinfo {author} {\bibfnamefont {A.}~\bibnamefont {Vishwanath}},\
  and\ \bibinfo {author} {\bibfnamefont {B.~I.}\ \bibnamefont {Halperin}},\
  }\bibfield  {title} {\bibinfo {title} {Topological order from disorder and
  the quantized {Hall} thermal metal: {Possible} applications to the $\nu$ =
  5/2 state},\ }\href {https://doi.org/10.1103/PhysRevB.98.045112} {\bibfield
  {journal} {\bibinfo  {journal} {Phys. Rev. B}\ }\textbf {\bibinfo {volume}
  {98}},\ \bibinfo {pages} {045112} (\bibinfo {year} {2018})}\BibitemShut
  {NoStop}%
\bibitem [{\citenamefont {Simon}(2018)}]{Simon2018}%
  \BibitemOpen
  \bibfield  {author} {\bibinfo {author} {\bibfnamefont {S.~H.}\ \bibnamefont
  {Simon}},\ }\bibfield  {title} {\bibinfo {title} {Interpretation of {Thermal}
  {Conductance} of the nu=5/2 {Edge}},\ }\href
  {https://doi.org/10.1103/PhysRevB.97.121406} {\bibfield  {journal} {\bibinfo
  {journal} {Phys. Rev. B}\ }\textbf {\bibinfo {volume} {97}},\ \bibinfo
  {pages} {121406} (\bibinfo {year} {2018})}\BibitemShut {NoStop}%
\bibitem [{\citenamefont {Ma}\ and\ \citenamefont
  {Feldman}(2019)}]{ma_partial_2019}%
  \BibitemOpen
  \bibfield  {author} {\bibinfo {author} {\bibfnamefont {K.~K.~W.}\
  \bibnamefont {Ma}}\ and\ \bibinfo {author} {\bibfnamefont {D.~E.}\
  \bibnamefont {Feldman}},\ }\bibfield  {title} {\bibinfo {title} {Partial
  equilibration of integer and fractional edge channels in the thermal quantum
  {Hall} effect},\ }\href {https://link.aps.org/doi/10.1103/PhysRevB.99.085309}
  {\bibfield  {journal} {\bibinfo  {journal} {Phys. Rev. B}\ }\textbf {\bibinfo
  {volume} {99}},\ \bibinfo {pages} {085309} (\bibinfo {year}
  {2019})}\BibitemShut {NoStop}%
\bibitem [{\citenamefont {Simon}\ and\ \citenamefont
  {Rosenow}(2019)}]{Simon2019}%
  \BibitemOpen
  \bibfield  {author} {\bibinfo {author} {\bibfnamefont {S.~H.}\ \bibnamefont
  {Simon}}\ and\ \bibinfo {author} {\bibfnamefont {B.}~\bibnamefont
  {Rosenow}},\ }\bibfield  {title} {\bibinfo {title} {Partial {Equilibration}
  of the {Anti}-{Pfaffian} edge due to {Majorana} {Disorder}},\ }\href
  {https://doi.org/10.1103/PhysRevLett.124.126801} {\bibfield  {journal}
  {\bibinfo  {journal} {Phys. Rev. Lett.}\ }\textbf {\bibinfo {volume} {124}},\
  \bibinfo {pages} {126801} (\bibinfo {year} {2019})}\BibitemShut {NoStop}%
\bibitem [{\citenamefont {Asasi}\ and\ \citenamefont
  {Mulligan}(2020)}]{Asasi2020}%
  \BibitemOpen
  \bibfield  {author} {\bibinfo {author} {\bibfnamefont {H.}~\bibnamefont
  {Asasi}}\ and\ \bibinfo {author} {\bibfnamefont {M.}~\bibnamefont
  {Mulligan}},\ }\bibfield  {title} {\bibinfo {title} {Partial equilibration of
  anti-{Pfaffian} edge modes at $\nu=5/2$},\ }\href
  {http://arxiv.org/abs/2004.04161} {\bibfield  {journal} {\bibinfo  {journal}
  {Phys. Rev. A}\ }\textbf {\bibinfo {volume} {102}},\ \bibinfo {pages}
  {205104} (\bibinfo {year} {2020})}\BibitemShut {NoStop}%
\bibitem [{\citenamefont {Balram}\ \emph {et~al.}(2018)\citenamefont {Balram},
  \citenamefont {Barkeshli},\ and\ \citenamefont
  {Rudner}}]{balram_parton_2018}%
  \BibitemOpen
  \bibfield  {author} {\bibinfo {author} {\bibfnamefont {A.~C.}\ \bibnamefont
  {Balram}}, \bibinfo {author} {\bibfnamefont {M.}~\bibnamefont {Barkeshli}},\
  and\ \bibinfo {author} {\bibfnamefont {M.~S.}\ \bibnamefont {Rudner}},\
  }\bibfield  {title} {\bibinfo {title} {Parton construction of a wave function
  in the anti-{Pfaffian} phase},\ }\href
  {https://link.aps.org/doi/10.1103/PhysRevB.98.035127} {\bibfield  {journal}
  {\bibinfo  {journal} {Phys. Rev. B}\ }\textbf {\bibinfo {volume} {98}},\
  \bibinfo {pages} {035127} (\bibinfo {year} {2018})}\BibitemShut {NoStop}%
\bibitem [{\citenamefont {Mishmash}\ \emph {et~al.}(2018)\citenamefont
  {Mishmash}, \citenamefont {Mross}, \citenamefont {Alicea},\ and\
  \citenamefont {Motrunich}}]{mishmash_numerical_2018}%
  \BibitemOpen
  \bibfield  {author} {\bibinfo {author} {\bibfnamefont {R.~V.}\ \bibnamefont
  {Mishmash}}, \bibinfo {author} {\bibfnamefont {D.~F.}\ \bibnamefont {Mross}},
  \bibinfo {author} {\bibfnamefont {J.}~\bibnamefont {Alicea}},\ and\ \bibinfo
  {author} {\bibfnamefont {O.~I.}\ \bibnamefont {Motrunich}},\ }\bibfield
  {title} {\bibinfo {title} {Numerical exploration of trial wave functions for
  the particle-hole-symmetric {Pfaffian}},\ }\href
  {https://doi.org/10.1103/PhysRevB.98.081107} {\bibfield  {journal} {\bibinfo
  {journal} {Phys. Rev. B}\ }\textbf {\bibinfo {volume} {98}},\ \bibinfo
  {pages} {081107} (\bibinfo {year} {2018})}\BibitemShut {NoStop}%
\bibitem [{\citenamefont {Pakrouski}(2021)}]{pakrouski_approximate_2021}%
  \BibitemOpen
  \bibfield  {author} {\bibinfo {author} {\bibfnamefont {K.}~\bibnamefont
  {Pakrouski}},\ }\bibfield  {title} {\bibinfo {title} {Approximate two-body
  generating {Hamiltonian} for the particle-hole {Pfaffian} wave function},\
  }\href {https://link.aps.org/doi/10.1103/PhysRevB.104.245306} {\bibfield
  {journal} {\bibinfo  {journal} {Phys. Rev. B}\ }\textbf {\bibinfo {volume}
  {104}},\ \bibinfo {pages} {245306} (\bibinfo {year} {2021})}\BibitemShut
  {NoStop}%
\bibitem [{\citenamefont {M{\"o}ller}\ and\ \citenamefont
  {Simon}(2008)}]{Moller2008}%
  \BibitemOpen
  \bibfield  {author} {\bibinfo {author} {\bibfnamefont {G.}~\bibnamefont
  {M{\"o}ller}}\ and\ \bibinfo {author} {\bibfnamefont {S.~H.}\ \bibnamefont
  {Simon}},\ }\bibfield  {title} {\bibinfo {title} {Paired composite-fermion
  wave functions},\ }\href@noop {} {\bibfield  {journal} {\bibinfo  {journal}
  {Phys. Rev. B}\ }\textbf {\bibinfo {volume} {77}},\ \bibinfo {pages} {075319}
  (\bibinfo {year} {2008})}\BibitemShut {NoStop}%
\bibitem [{\citenamefont {Sharma}\ \emph {et~al.}(2021)\citenamefont {Sharma},
  \citenamefont {Pu},\ and\ \citenamefont {Jain}}]{Sharma2021}%
  \BibitemOpen
  \bibfield  {author} {\bibinfo {author} {\bibfnamefont {A.}~\bibnamefont
  {Sharma}}, \bibinfo {author} {\bibfnamefont {S.}~\bibnamefont {Pu}},\ and\
  \bibinfo {author} {\bibfnamefont {J.~K.}\ \bibnamefont {Jain}},\ }\bibfield
  {title} {\bibinfo {title} {Bardeen-{Cooper}-{Schrieffer} pairing of composite
  fermions},\ }\href@noop {} {\bibfield  {journal} {\bibinfo  {journal} {Phys.
  Rev. B}\ }\textbf {\bibinfo {volume} {104}},\ \bibinfo {pages} {205303}
  (\bibinfo {year} {2021})}\BibitemShut {NoStop}%
\bibitem [{\citenamefont {Bardeen}\ \emph {et~al.}(1957)\citenamefont
  {Bardeen}, \citenamefont {Cooper},\ and\ \citenamefont
  {Schrieffer}}]{bardeen_theory_1957}%
  \BibitemOpen
  \bibfield  {author} {\bibinfo {author} {\bibfnamefont {J.}~\bibnamefont
  {Bardeen}}, \bibinfo {author} {\bibfnamefont {L.~N.}\ \bibnamefont
  {Cooper}},\ and\ \bibinfo {author} {\bibfnamefont {J.~R.}\ \bibnamefont
  {Schrieffer}},\ }\bibfield  {title} {\bibinfo {title} {Theory of
  {Superconductivity}},\ }\href
  {https://link.aps.org/doi/10.1103/PhysRev.108.1175} {\bibfield  {journal}
  {\bibinfo  {journal} {Phys. Rev.}\ }\textbf {\bibinfo {volume} {108}},\
  \bibinfo {pages} {1175} (\bibinfo {year} {1957})}\BibitemShut {NoStop}%
\bibitem [{\citenamefont {Yutushui}\ and\ \citenamefont
  {Mross}(2020)}]{Yutushui2020}%
  \BibitemOpen
  \bibfield  {author} {\bibinfo {author} {\bibfnamefont {M.}~\bibnamefont
  {Yutushui}}\ and\ \bibinfo {author} {\bibfnamefont {D.~F.}\ \bibnamefont
  {Mross}},\ }\bibfield  {title} {\bibinfo {title} {Large-scale simulations of
  particle-hole-symmetric {Pfaffian} trial wave functions},\ }\href
  {https://doi.org/10.1103/PhysRevB.102.195153} {\bibfield  {journal} {\bibinfo
   {journal} {Phys. Rev. B}\ }\textbf {\bibinfo {volume} {102}},\ \bibinfo
  {pages} {195153} (\bibinfo {year} {2020})}\BibitemShut {NoStop}%
\bibitem [{\citenamefont {Wu}\ and\ \citenamefont {Yang}(1976)}]{Wu1976}%
  \BibitemOpen
  \bibfield  {author} {\bibinfo {author} {\bibfnamefont {T.~T.}\ \bibnamefont
  {Wu}}\ and\ \bibinfo {author} {\bibfnamefont {C.~N.}\ \bibnamefont {Yang}},\
  }\bibfield  {title} {\bibinfo {title} {Dirac monopole without strings:
  {Monopole} harmonics},\ }\href {https://doi.org/10.1016/0550-3213(76)90143-7}
  {\bibfield  {journal} {\bibinfo  {journal} {Nuc. Phys. B}\ }\textbf {\bibinfo
  {volume} {107}},\ \bibinfo {pages} {365} (\bibinfo {year}
  {1976})}\BibitemShut {NoStop}%
\bibitem [{\citenamefont {Jain}\ and\ \citenamefont
  {Kamilla}(1997)}]{Jain1997}%
  \BibitemOpen
  \bibfield  {author} {\bibinfo {author} {\bibfnamefont {J.~K.}\ \bibnamefont
  {Jain}}\ and\ \bibinfo {author} {\bibfnamefont {R.~K.}\ \bibnamefont
  {Kamilla}},\ }\bibfield  {title} {\bibinfo {title} {Composite fermions in the
  {Hilbert} space of the lowest electronic {Landau} level},\ }\href
  {https://doi.org/10.1142/S0217979297001301} {\bibfield  {journal} {\bibinfo
  {journal} {Int. J. Mod. Phys. B}\ }\textbf {\bibinfo {volume} {11}},\
  \bibinfo {pages} {2621} (\bibinfo {year} {1997})}\BibitemShut {NoStop}%
\bibitem [{\citenamefont {M{\"o}ller}(2006)}]{mollerPhDThesis2006}%
  \BibitemOpen
  \bibfield  {author} {\bibinfo {author} {\bibfnamefont {G.}~\bibnamefont
  {M{\"o}ller}},\ }\emph {\bibinfo {title} {Dynamically Reduced Spaces in
  Condensed Matter Physics:{{Quantum Hall}} Bilayers, Dimensional Reduction,
  and Magnetic Spin Systems}},\ \href
  {https://theses.hal.science/tel-00121765/en/} {Ph.D. thesis},\ \bibinfo
  {school} {Universit{\'e} Paris XI}, \bibinfo {address} {{Orsay}} (\bibinfo
  {year} {2006})\BibitemShut {NoStop}%
\bibitem [{\citenamefont {M{\"o}ller}\ \emph {et~al.}(2008)\citenamefont
  {M{\"o}ller}, \citenamefont {Simon},\ and\ \citenamefont
  {Rezayi}}]{Moller2008BilayerPRL}%
  \BibitemOpen
  \bibfield  {author} {\bibinfo {author} {\bibfnamefont {G.}~\bibnamefont
  {M{\"o}ller}}, \bibinfo {author} {\bibfnamefont {S.~H.}\ \bibnamefont
  {Simon}},\ and\ \bibinfo {author} {\bibfnamefont {E.~H.}\ \bibnamefont
  {Rezayi}},\ }\bibfield  {title} {\bibinfo {title} {Paired composite fermion
  phase of quantum hall bilayers at {{$\nu$}}=(1)/(2)+(1)/(2)},\ }\href
  {https://doi.org/10.1103/PhysRevLett.101.176803} {\bibfield  {journal}
  {\bibinfo  {journal} {Phys. Rev. Lett.}\ }\textbf {\bibinfo {volume} {101}},\
  \bibinfo {pages} {176803} (\bibinfo {year} {2008})}\BibitemShut {NoStop}%
\bibitem [{\citenamefont {M{\"o}ller}\ \emph {et~al.}(2009)\citenamefont
  {M{\"o}ller}, \citenamefont {Simon},\ and\ \citenamefont
  {Rezayi}}]{Moller2009BilayerPRB}%
  \BibitemOpen
  \bibfield  {author} {\bibinfo {author} {\bibfnamefont {G.}~\bibnamefont
  {M{\"o}ller}}, \bibinfo {author} {\bibfnamefont {S.~H.}\ \bibnamefont
  {Simon}},\ and\ \bibinfo {author} {\bibfnamefont {E.~H.}\ \bibnamefont
  {Rezayi}},\ }\bibfield  {title} {\bibinfo {title} {Trial wave functions for
  {{$\nu$}}=(1)/(2)+(1)/(2) quantum {{Hall}} bilayers},\ }\href
  {https://doi.org/10.1103/PhysRevB.79.125106} {\bibfield  {journal} {\bibinfo
  {journal} {Phys. Rev. B}\ }\textbf {\bibinfo {volume} {79}},\ \bibinfo
  {pages} {125106} (\bibinfo {year} {2009})}\BibitemShut {NoStop}%
\bibitem [{\citenamefont {Wu}\ and\ \citenamefont {Yang}(1977)}]{Wu1977}%
  \BibitemOpen
  \bibfield  {author} {\bibinfo {author} {\bibfnamefont {T.~T.}\ \bibnamefont
  {Wu}}\ and\ \bibinfo {author} {\bibfnamefont {C.~N.}\ \bibnamefont {Yang}},\
  }\bibfield  {title} {\bibinfo {title} {Some properties of monopole
  harmonics},\ }\href@noop {} {\bibfield  {journal} {\bibinfo  {journal} {Phys.
  Rev. D}\ }\textbf {\bibinfo {volume} {16}},\ \bibinfo {pages} {1018}
  (\bibinfo {year} {1977})}\BibitemShut {NoStop}%
\bibitem [{Note1()}]{Note1}%
  \BibitemOpen
  \bibinfo {note} {This can be seen using stereographic coordinates with $z_i =
  u_i/v_i$ which then gives $(u_iv_j-v_iu_j)^2q =
  v_i^{2q}v_j^{2q}(z_i-z_j)^{2q}$. Hence, as we bring particle $i$ around
  particle $j$ the phase of the wave function will wind $2q$ times in the
  anti-clockwise direction.}\BibitemShut {Stop}%
\bibitem [{\citenamefont {Wooten}(2013)}]{Wooten2013}%
  \BibitemOpen
  \bibfield  {author} {\bibinfo {author} {\bibfnamefont {R.~E.}\ \bibnamefont
  {Wooten}},\ }\emph {\bibinfo {title} {Haldane pseudopotentials and {Landau}
  level mixing in the quantum {Hall} effect}},\ \href
  {https://trace.tennessee.edu/utk_graddiss/1796} {Ph.D. thesis} (\bibinfo
  {year} {2013})\BibitemShut {NoStop}%
\bibitem [{\citenamefont {Morf}\ and\ \citenamefont
  {Halperin}(1987)}]{morf_monte_1987}%
  \BibitemOpen
  \bibfield  {author} {\bibinfo {author} {\bibfnamefont {R.}~\bibnamefont
  {Morf}}\ and\ \bibinfo {author} {\bibfnamefont {B.~I.}\ \bibnamefont
  {Halperin}},\ }\bibfield  {title} {\bibinfo {title} {Monte {Carlo} evaluation
  of trial wavefunctions for the fractional quantized {Hall} effect:
  {Spherical} geometry},\ }\href {https://doi.org/10.1007/BF01304256}
  {\bibfield  {journal} {\bibinfo  {journal} {Z. Physik B}\ }\textbf {\bibinfo
  {volume} {68}},\ \bibinfo {pages} {391} (\bibinfo {year} {1987})}\BibitemShut
  {NoStop}%
\bibitem [{\citenamefont {M{\"o}ller}\ and\ \citenamefont
  {Simon}(2005)}]{moller_composite_2005}%
  \BibitemOpen
  \bibfield  {author} {\bibinfo {author} {\bibfnamefont {G.}~\bibnamefont
  {M{\"o}ller}}\ and\ \bibinfo {author} {\bibfnamefont {S.~H.}\ \bibnamefont
  {Simon}},\ }\bibfield  {title} {\bibinfo {title} {Composite fermions in a
  negative effective magnetic field: {A} {Monte} {Carlo} study},\ }\href
  {https://link.aps.org/doi/10.1103/PhysRevB.72.045344} {\bibfield  {journal}
  {\bibinfo  {journal} {Phys. Rev. B}\ }\textbf {\bibinfo {volume} {72}},\
  \bibinfo {pages} {045344} (\bibinfo {year} {2005})}\BibitemShut {NoStop}%
\bibitem [{Note2()}]{Note2}%
  \BibitemOpen
  \bibinfo {note} {I.e. fitting quadratic polynomials can be much less stable
  than fitting a linear function}\BibitemShut {NoStop}%
\bibitem [{\citenamefont {Haldane}(2023)}]{haldane2023incompressible}%
  \BibitemOpen
  \bibfield  {author} {\bibinfo {author} {\bibfnamefont {F.}~\bibnamefont
  {Haldane}},\ }\bibfield  {title} {\bibinfo {title} {Incompressible quantum
  hall fluids as electric quadrupole fluids},\ }\href
  {https://arxiv.org/abs/2302.12472} {\bibfield  {journal} {\bibinfo  {journal}
  {arXiv preprint arXiv:2302.12472}\ } (\bibinfo {year} {2023})}\BibitemShut
  {NoStop}%
\bibitem [{\citenamefont {von~der Linden}(1992)}]{von_der_linden_quantum_1992}%
  \BibitemOpen
  \bibfield  {author} {\bibinfo {author} {\bibfnamefont {W.}~\bibnamefont
  {von~der Linden}},\ }\bibfield  {title} {\bibinfo {title} {A quantum {Monte}
  {Carlo} approach to many-body physics},\ }\href
  {https://www.sciencedirect.com/science/article/pii/037015739290029Y}
  {\bibfield  {journal} {\bibinfo  {journal} {Phys. Rep.}\ }\textbf {\bibinfo
  {volume} {220}},\ \bibinfo {pages} {53} (\bibinfo {year} {1992})}\BibitemShut
  {NoStop}%
\bibitem [{\citenamefont {Sorella}\ \emph {et~al.}(2007)\citenamefont
  {Sorella}, \citenamefont {Casula},\ and\ \citenamefont
  {Rocca}}]{Sorella2007}%
  \BibitemOpen
  \bibfield  {author} {\bibinfo {author} {\bibfnamefont {S.}~\bibnamefont
  {Sorella}}, \bibinfo {author} {\bibfnamefont {M.}~\bibnamefont {Casula}},\
  and\ \bibinfo {author} {\bibfnamefont {D.}~\bibnamefont {Rocca}},\ }\bibfield
   {title} {\bibinfo {title} {Weak binding between two aromatic rings:
  {Feeling} the van der {Waals} attraction by quantum {Monte} {Carlo}
  methods},\ }\href@noop {} {\bibfield  {journal} {\bibinfo  {journal} {J.
  Chem. Phys.}\ }\textbf {\bibinfo {volume} {127}} (\bibinfo {year}
  {2007})}\BibitemShut {NoStop}%
\bibitem [{\citenamefont {Kingma}\ and\ \citenamefont {Ba}(2015)}]{Kingma2015}%
  \BibitemOpen
  \bibfield  {author} {\bibinfo {author} {\bibfnamefont {D.~P.}\ \bibnamefont
  {Kingma}}\ and\ \bibinfo {author} {\bibfnamefont {J.~L.}\ \bibnamefont
  {Ba}},\ }\bibfield  {title} {\bibinfo {title} {Adam: {A} method for
  stochastic optimization},\ }in\ \href {https://arxiv.org/abs/1412.6980v9}
  {\emph {\bibinfo {booktitle} {3rd {International} {Conference} on {Learning}
  {Representations}, {ICLR} 2015 - {Conference} {Track} {Proceedings}}}}\
  (\bibinfo  {publisher} {ICLR},\ \bibinfo {year} {2015})\BibitemShut {NoStop}%
\bibitem [{\citenamefont {Carleo}\ and\ \citenamefont
  {Troyer}(2017)}]{Carleo2017}%
  \BibitemOpen
  \bibfield  {author} {\bibinfo {author} {\bibfnamefont {G.}~\bibnamefont
  {Carleo}}\ and\ \bibinfo {author} {\bibfnamefont {M.}~\bibnamefont
  {Troyer}},\ }\bibfield  {title} {\bibinfo {title} {Solving the quantum
  many-body problem with artificial neural networks},\ }\href
  {https://www.science.org} {\bibfield  {journal} {\bibinfo  {journal}
  {Science}\ }\textbf {\bibinfo {volume} {355}},\ \bibinfo {pages} {602}
  (\bibinfo {year} {2017})}\BibitemShut {NoStop}%
\end{thebibliography}%

\end{document}